\documentclass[preprint,groupedaddress,floatfix,%
nofootinbib]{revtex4}

\usepackage{euscript}
\usepackage{dcolumn}
\usepackage{graphicx}
\usepackage{epsfig}
\usepackage{hyperref}
\usepackage{graphicx}
\usepackage{dcolumn}
\usepackage{longtable}
\usepackage{times}
\usepackage{amsmath,amssymb,bm}
\usepackage[english]{babel}
\usepackage[latin1]{inputenc}
\usepackage{times}
\usepackage[T1]{fontenc}
\usepackage{color}
\usepackage{fancybox}
\usepackage{sidecap}
\usepackage{graphicx}
\usepackage{epsfig}
\usepackage{psfrag}
\usepackage{bbm}
\usepackage[english]{babel}
\usepackage[latin1]{inputenc}
\usepackage{times}
\usepackage[T1]{fontenc}
\usepackage{color}
\usepackage{fancybox}
\usepackage{sidecap}
\usepackage{psfrag}
\usepackage{bbm}
\usepackage{wasysym}
\usepackage{slashed}
\usepackage{tikz}
\usepackage{euscript}

\usepackage{algorithm}
\usepackage{algpseudocode}

\newcommand{\mathe}{\mathrm{e}}

\begin{document}

\title{More on greedy construction heuristics for the MAX-CUT
problem}
\author{Jianan Wang}
\author{Chuixiong Wu}
\author{Fen Zuo\footnote{Email: \textsf{zuofen@miqroera.com}}}
\affiliation{Hefei MiQro Era Digital Technology Co. Ltd., Hefei, China}

\begin{abstract}
A cut of a graph can be represented in many different ways. Here we propose to represent a cut through a ``relation tree'', which is a spanning tree with signed edges. We show that this picture helps to classify the main greedy heuristics for the maximum cut problem, in analogy with the minimum spanning tree problem. Namely, all versions of the Sahni-Gonzalez~(SG) algorithms could be classified as the Prim class, while various Edge-Contraction~(EC) algorithms are of the Kruskal class. We further elucidate the relation of this framework to the stabilizer formalism in quantum computing, and point out that the recently proposed \textit{ADAPT-Clifford} algorithm is a reformulation of a refined version of the SG algorithm, SG3. Numerical performance of the typical algorithms from the two classes are studied with various kinds of graphs. It turns out that, the Prim-class algorithms perform better for general dense graphs, and the Kruskal-class algorithms performs better when the graphs are sparse enough.

\end{abstract}
 \maketitle

\tableofcontents

\section{Introduction}

 Finding the maximum-weight cut for a weighted undirected graph, MAX-CUT for short, is a celebrated NP-hard problem. This hardness persists even in the unweighted case. Thus we have to turn to approximate algorithms. Up to now, the best theoretical approximation ratio is provided by the Goemans-Williamson~(GW) algorithm~\cite{GW-1995}, which is based on semi-definite programming~(SDP). However, as emphasized in~\cite{Kahruman-2007}, the time and space requirements for SDP approaches limit their applicability in practice. Therefore, development of fast and simple heuristics would still be of practical importance.

There are two main classes of greedy construction heuristics for MAX-CUT: the Sahni-Gonzalez~(SG) approach and the Edge-Contraction~(EC) approach. The original SG algorithm was proposed in 1976~\cite{SG-1976}, which is perhaps the first algorithm for MAX-CUT. While the original algorithm deals with the vertices in a fixed order, some deliberate criterions for vertex-selecting were made in \cite{Kahruman-2007}, which greatly improve the performance. Ref.~\cite{Kahruman-2007} also proposed a new construction heuristic: Edge Contraction. This bears remarkable similarity to the random contraction algorithm for the Minimum Cut problem~(MIN-CUT)~\cite{KS-1993}. While the original EC is a ``worst-out'' approach, in ref.~\cite{Hassin-2021} the corresponding ``best-in'' approach is developed, and named ``Differencing Edge Contraction''~(DEC). More interestingly, recently we find that these two could be unified into a single ``best-in-worst-out'' algorithm~\cite{SAIII}, which could be properly named as ``Signed Edge Contraction''~(SEC). Are these two kinds of heuristics related to each other? One may think that the two are so different that no relation of any kind could exist. However, we show in this paper that they could actually be unified into a single framework. This is achieved by representing the cut with a ``relation tree'', then the SG and EC approaches are simply different ways to generate the tree.

In fact, the concept of ``relation tree'' has already appeared in our implementation of the SEC heuristic~\cite{SAIII}. Even more importantly, the SEC heuristic is not developed by generalizing the EC and DEC approach, but from a completely different way of thinking. It originates from a very disparate area, quantum computing. More precisely, it is a manifestation of the stabilizer formalism in graph theory. This is why SEC is originally called ``stabilizer heuristic'', and the relation tree called ``stabilizer tree'' in~\cite{SAIII}. We elucidate the relation between the graph formulation and the stabilizer formulation here. With this relation clarified, we then examine another recently proposed algorithm based on the stabilizer formalism, \textit{ADAPT-Clifford}~\cite{ADAPT-Clifford}, and show that it is just a reformulation of a refined version of SG.

In the remaining part we analyze the performance of these algorithms with various graphs. These include complete graphs with positive weights or signed weights, randomly generated (un)weighted $k$-regular graphs, and (un)weighted Erd\"{o}s-R\'{e}nyi graphs $G(n,p)$. The results from the GW algorithm are taken as references, when they are available. We are mainly interested in two questions: when would these greedy algorithms be competitive to GW? and which of them would perform best? In seeking the answers to these questions, we discover some new features of these algorithms not noticed before.

The arrangement of the paper is as follows. In section II we introduce the concept of a relation tree, and use this concept in section III to classify the usual greedy algorithms for MAX-CUT. In section IV we introduce the stabilizer formalism, and revisit the previous algorithms and the recently proposed \textit{ADAPT-Clifford}. The performance of these algorithms is analyzed in section V in detail. In the final section we give a short summary.

\section{Relation Tree}

Given an undirected weighted graph $G=(V, E, w)$, an edge set $C\subset E$ is called a cut if for some $A\subset V$,
\begin{equation}
C=\delta(A)\equiv\{(i,j)\in E|i\in A,j\in V\setminus A\}.
\end{equation}
So in order to specify the cut $C$, we just need to specify $A$. A common way to do this is to use the characteristic vector $x$ of $A$. Let $V=\{1,...,n\}$, then $x$ is an $n$-dimensional vector with its elements taking binary values:
\begin{eqnarray}
x_i=1, &&\text{if} \quad i\in A\\
x_i=0, &&\text{otherwise}.
\end{eqnarray}
Notice that $x$ and its negation $\neg x$ (with elements $1-x_i$) give rise to the same cut. With this characterization, the weight of a cut $C$ can be expressed as
\begin{equation}
w(x)=\sum_{i,j} w_{ij} x_i(1-x_j).
\end{equation}
And the maximum cut problem requires to find a cut with the maximum cut weight.

Of course the binary-valued vector is not the unique choice to represent the cut. It has been known for a long time that MAX-CUT is equivalent to the Ising model in physics~\cite{BGJR-1988}. Ising model is naturally expressed with spin values $z_i\in \{-1, +1\}$. Therefore, we could also use a spin-valued vector to characterize the cut:
\begin{eqnarray}
z_i=-1, &&\text{if} \quad i\in A\\
z_i=+1, &&\text{otherwise}.
\end{eqnarray}
That is, we could turn the vector $x$ into $z$ with the relation:
\begin{equation}
z_i=1-2x_i.
\end{equation}
Then the negation of $x$, $\neg x$, corresponds to $-z$. Of course $-z$ also represents the same cut as $z$. With the spin values, we may express the cut weight as
\begin{equation}
w(z)=\frac{1}{2}\sum_{i<j} w_{ij} (1-z_i z_j). \label{eq:wz}
\end{equation}
Interestingly, the celebrated GW algorithm is based on a relaxation of this expression~\cite{GW-1995}. Moreover, the spin representation is very convenient to distinguish the edges inside and outside the cut. We just need to evaluate the product of the spin values of the vertices incident with the edge. The edge $(i,j)\in E$ is in the cut if and only if $z_iz_j=-1$.

We could also reverse the process: in order to obtain all the spin values, we could first seek the relations between them. The total number of relations is certainly much larger than the number of vertices. For $n$ vertices we would have in total $n(n-1)/2$ relations. But they are correlated; only $n-1$ of them are independent. These must have support on the edges of a spanning tree. Therefore we have the following definition:\\
\textbf{Definition 1}:\\
For an undirected graph $G=(V, E)$ with $|V|=n$, a \textbf{relation tree} $H=(V, T, s)$ is a spanning tree $(V,T)$ of $K_n$ together with an assignment of relations on its edges:
\begin{equation}
s: T\to \{-1,+1\}.
\end{equation}
 When the edge is assigned a ``+1'' value, we call it ``positive'', and ``negative'' otherwise. The definition looks a little weird: the relation tree needs not be a spanning tree of $G$ itself. However, we could engineer it using the information of $G$. It is not difficult to prove that such a relation tree generates a unique cut. To explicitly obtain the cut, one could apply the graph-scanning algorithm for $H$~\cite{SAIII}. In order for the cut to be nontrivial, the values $s$ should not be all positive. Notice that the relation tree is different from the cut tree in the famous Gomory-Hu algorithm~\cite{GH-1961} for the Multi-terminal MIN-CUT problem. Also notice that the relation on an edge should not be considered as a direction.

Let us give an example of this. For the following graph,
\begin{figure}[H]
\centering
	\includegraphics[width=0.6\textwidth]{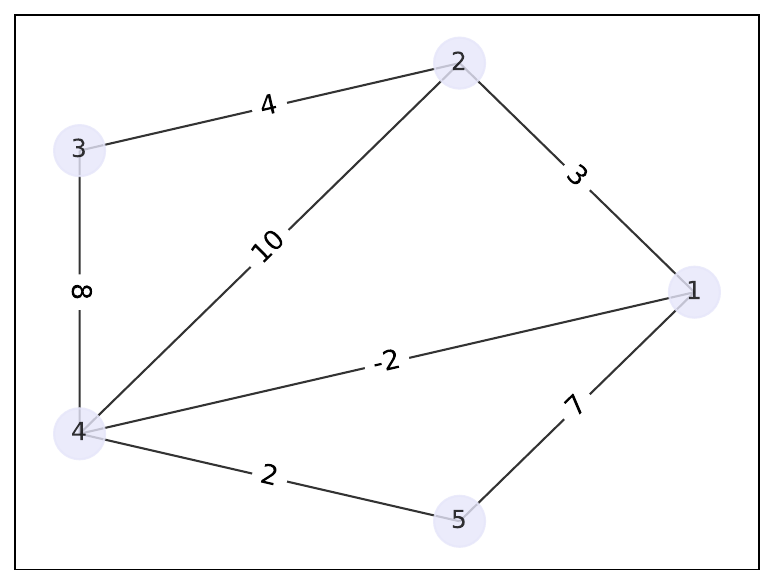}
\caption{\it A weighted graph.}\label{fig:G5}
\end{figure}
the maximum cut is given by $\{1,4\}$ and $\{2,3,5\}$. We may express such a cut with the relation tree:
\begin{figure}[H]
\centering
	\includegraphics[width=0.6\textwidth]{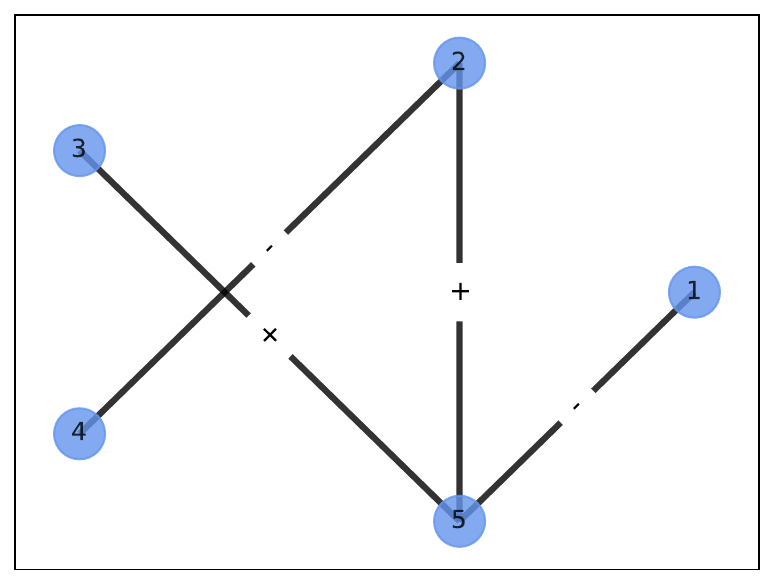}
\caption{\it A relation tree of the graph in Fig.~\ref{fig:G5}.}\label{fig:G5-Tree}
\end{figure}
The relation tree gives a nice visualization of the cut. Of course such a tree is not unique. A specific cut may be represented by many different but equivalent relation trees.

\section{Tree Algorithms}

 So what is the usefulness of such a concept, a relation tree? Once the focus is put on the relations trees, many nice properties of spanning trees could be employed. In particular, we could use the wellknown algorithms for the minimum spanning tree~(MST) as a guide to classify the MAX-CUT algorithms. As we know, MST has two celebrated algorithms: the Prim algorithm~\cite{Prim-1957} and the Kruskal algorithm~\cite{Kruskal-1956}. Roughly speaking, Prim algorithm is vertex-oriented, while Kruskal algorithm is edge-oriented. Accordingly, the MAX-CUT algorithms based on the relation tree can also be decomposed into two classes: the Prim class and the Kruskal class. We will discuss them one by one.

 \subsection{Prim class}
For the sake of clarity, we give the pseudo-code of the original Prim algorithm here:

\begin{algorithm}[H]
\caption{Prim's Algorithm for MST}\label{algo:Prim}
\begin{algorithmic}
\State \textbf{Input:} A weighted graph $G=(V,E,w)$ with $w: E\to \mathbb{R}$
\State \textbf{Output:} A minimum spanning tree $H=(V,T)$
\State Initialize $H=(V(H),T)$ as $(\{r\},\varnothing)$ with $r$ a randomly chosen vertex
\While {$H$ is not a spanning tree}
        \State Adding to $T$ a minimum-weight edge from $\delta(V(H))$.
\EndWhile
\end{algorithmic}
\end{algorithm}

As we have mentioned, the essence of the Prim algorithm is vertex-orientation. It starts with a vertex, and at each step adds a new vertex through a minimum-weight edge. Meanwhile, $H$ is a tree from the beginning to the end. This reminds us of the SG algorithm~\cite{SG-1976}, which we rewrite here:

\begin{algorithm}[H]
\caption{SG Algorithm for MAX-CUT}\label{algo:SG}
\begin{algorithmic}
\State \textbf{Input:} A weighted graph $G=(V,E,w)$ with $w: E\to \mathbb{R}$
\State \textbf{Output:} A partition $V=V_1\uplus V_2$ and the cut weight $w(V_1,V_2)$
\State Order the vertex set as $V=\{1,2,...,n\}$
\State Initialize $V_1=\{1\}, V_2=\{2\}$
\State $w(V_1,V_2) \gets w_{12}$
\For {$i=3:n$}
     \State $w(i,V_1)=\sum_{(i,j)\in E,j\in V_1}w(i,j)$
     \State $w(i,V_2)=\sum_{(i,j)\in E,j\in V_2}w(i,j)$
     \If {$w(i,V_1)> w(i,V_2)$}
         \State $V_2\gets V_2\cup \{i\}$
     \Else
         \State $V_1\gets V_1\cup \{i\}$
     \EndIf
     \State $w(V_1,V_2)\gets w(V_1,V_2)+\max\{w(i,V_1),w(i,V_2)\}$
\EndFor
\end{algorithmic}
\end{algorithm}

Here the first two vertices are always assigned to different sets, because in the original SG formulation all the weights are assumed to be non-negative. If we allow the weights to be negative, such an assignment may not be proper. It is not difficult to see that the SG algorithm is actually building a relation tree step by step: first choose edge $(1,2)$ as a negative leaf, then connect the other vertices to vertex $1$ or $2$ with positive edges. At each step, the choice between vertex $1$ and $2$ is made to maximize the current cut weight. For example, applying SG to Fig.~\ref{fig:G5}, we get the following relation tree:

\begin{figure}[H]
\centering
	\includegraphics[width=0.6\textwidth]{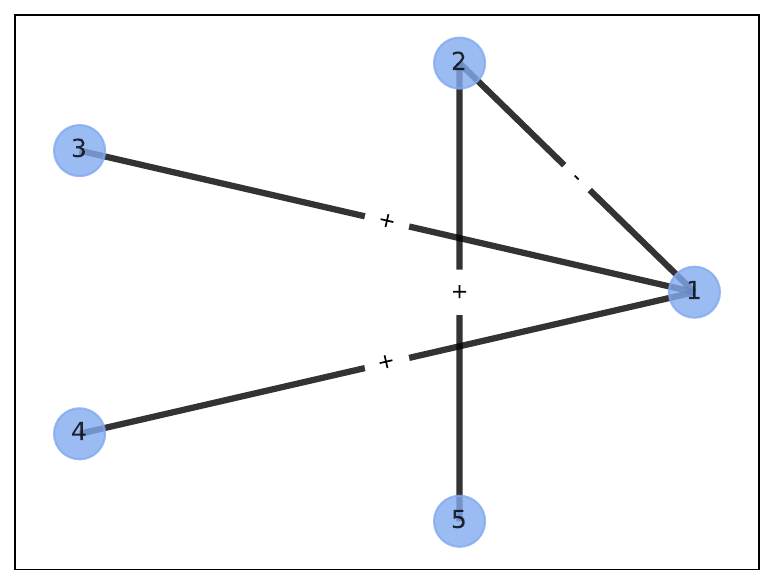}
\caption{\it Relation tree obtained by applying SG to the graph in Fig.~\ref{fig:G5}.}\label{fig:SG-Tree}
\end{figure}

However, SG is not really a Prim-type algorithm: it is not as ``greedy'' as the Prim algorithm! This is because in SG all the vertices are examined in a fixed order, which makes it very efficient but maybe not so efficacious. Indeed, if the graph has $m$ edges, the complexity of SG is $O(n+m)$~\cite{SG-1976}, which could even be lower than that of the Prim algorithm, which is at best $O(n^2)$.

To make SG more Prim-like, we could discard the ordering of the vertices, and add some criterion for the selection of the vertex at each step. For the initial vertex, it seems that no vertex will have priority and we have to randomly select one as the starting point. This is exactly what the Prim algorithm begins with. Alternatively, one could eliminate the randomness by choosing the maximum weighted edge first. Since this weight must be positive (otherwise the problem would be easy to solve by reducing it to MIN-CUT), one includes this edge in the cut. Warning here: this may not always be a good initialization. As for the criterion of vertex-selecting in the next steps, we have three different choices, namely ``best-in'', ``worst-out'', and ``best-in-worst-out''. Correspondingly, we have three modified versions of SG, SG1, SG2 and SG3 respectively~\cite{Kahruman-2007}. See also ~\cite{Bian-2019} for some related discussion. The pseudo-code of SG1 reads:

\begin{algorithm}[H]
\caption{SG1 Algorithm for MAX-CUT}\label{algo:SG1}
\begin{algorithmic}
\State \textbf{Input:} A weighted graph $G=(V,E,w)$ with $w: E\to \mathbb{R}$
\State \textbf{Output:} A partition $V=V_1\uplus V_2$ and the cut weight $w(V_1,V_2)$
\State Pick the maximum weighted edge $(i_1,i_2)$
\State $V_1\gets \{i_1\}, V_2\gets \{i_2\}$
\State $w(V_1,V_2) \gets w_{i_1i_2}$
\State $V'\gets V\setminus \{i_1,i_2\}$
\For {$j=1:n-2$}
     \For {$i\in V'$}
          \State $w(i,V_1)=\sum_{(i,j)\in E,j\in V_1}w(i,j)$
          \State $w(i,V_2)=\sum_{(i,j)\in E,j\in V_2}w(i,j)$
          \State $\textrm{score}(i)= \max\{w(i,V_1),w(i,V_2)\}$
     \EndFor
     \State
     $i^*\gets \arg \max_{i\in V'} {\textrm{score}(i)}$
         \If {$w(i^*,V_1)> w(i^*,V_2)$}
              \State $V_2\gets V_2\cup \{i^*\}$
         \Else
              \State $V_1\gets V_1\cup \{i^*\}$
         \EndIf
     \State $V'\gets V'\setminus \{i^*\}$
     \State $w(V_1,V_2)\gets w(V_1,V_2)+\max\{w(i^*,V_1),w(i^*,V_2)\}$
\EndFor
\end{algorithmic}
\end{algorithm}
Again we are actually building a relation tree here: first find the maximum weighted edge $(i_1,i_2)$ and set it as negative; then at each step add a new positive leaf to $i_1$ or $i_2$, depending on the benefit of the cut weight. Applying SG1 to  Fig.~\ref{fig:G5}, we get the following relation tree:\begin{figure}[H]
\centering
	\includegraphics[width=0.6\textwidth]{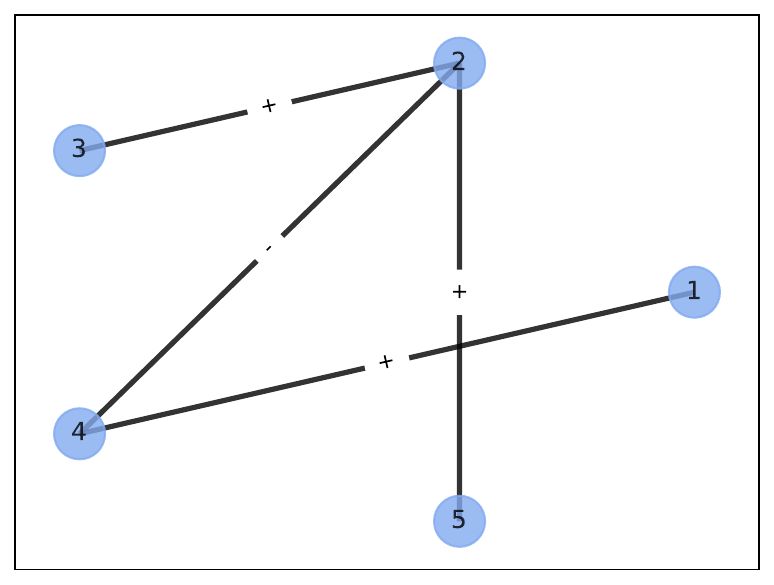}
\caption{\it Relation tree obtained by applying SG1 to the graph in Fig.~\ref{fig:G5}.}\label{fig:SG1-Tree}
\end{figure}

Alternatively, SG2 uses a ``worst-out'' criterion:
\begin{eqnarray}
 \textrm{\textrm{score}}(i)&=& \min\{w(i,V_1),w(i,V_2)\},\\
  i^*&\gets& \arg \min_{i\in V'} {\textrm{score}(i)}.\
 \end{eqnarray}
 And SG3 is a clever combination of them, and uses a ``best-in-worst-out'' criterion:
\begin{eqnarray}
 \textrm{score}(i)&=& |w(i,V_1)-w(i,V_2)|,\\
 i^*&\gets& \arg \max_{i\in V'} {\textrm{score}(i)}.
 \end{eqnarray}

All three algorithms have time complexity $O(n^2)$, as pointed out in~\cite{Hassin-2021}.

\subsubsection{Approximation Ratio}

The approximation ratio is usually defined with respect to the optimal cut weight $w_{opt}$. If all the edge weights are nonnegative, $w_{opt}$ is bounded from above by the total weight $w_{tot}$:
\begin{equation}
w_{opt}\le w_{tot} \equiv \sum_{i<j} w_{ij} .
\end{equation}
So if the optimal cut is not known in advance, we could use $w_{tot}$ instead to define the approximation ratio:
\begin{equation}
R_\mathcal{A}\equiv w_\mathcal{A}/w_{tot},
\end{equation}
where $w_\mathcal{A}$ is the cut weight obtained by algorithm $\mathcal {A}$. With such a definition the approximation ratio can not be strictly greater than $1/2$ theoretically, because there exist graphs whose optimal cut value approaches $w_{tot}/2$ from above.

We will show that all the four versions of SG algorithms have an approximation ratio $1/2$ in the above sense. Since the vertex set in the original SG algorithm can actually be given any order, and SG1-3 differ from SG only by vertex ordering, it suffices to prove this for SG itself. For this we define the current total edge weight as $w_0(V_1,V_2)$, and initiate it as $w_{12}$. We assume $w_{12}\ge0$, so initially we have
\begin{equation}
w(V_1,V_2)=w_{12}\ge \frac{w_{12}}{2}=\frac{1}{2} w_0(V_1,V_2).
\end{equation}
At each step the current total weight is updated as
\begin{equation}
w_0(V_1,V_2) \gets w_0(V_1,V_2)+w(i,V_1)+w(i,V_2).
\end{equation}
Since
\begin{equation}
\max\{w(i,V_1),w(i,V_2)\}\ge \frac{1}{2} (w(i,V_1)+w(i,V_2)),
\end{equation}
we have at every step
\begin{equation}
w(V_1,V_2)\ge \frac{1}{2} w_0(V_1,V_2).
\end{equation}
When finally we get the cut $V=V_1\uplus V_2$, we have $w_0(V_1,V_2)=w_{tot}$, and thus
\begin{equation}
R_{SG}= \frac{1}{2}.
\end{equation}
Numerically, it is shown in~\cite{Kahruman-2007} that SG3 performs much better than SG. We will discuss this in more detail later.

\subsection{Kruskal class}
Now we turn to the Kruskal-class algorithms. First we repeat the original Kruskal algorithm for MST:

\begin{algorithm}[H]
\caption{Kruskal's Algorithm for MST}\label{algo:Prim}
\begin{algorithmic}
\State \textbf{Input:} A weighted graph $G=(V,E,w)$ with $w: E\to \mathbb{R}$
\State \textbf{Output:} A minimum spanning tree $H=(V,F)$
\State Keep a spanning forest $H=(V,F)$ of $G$, with $F=\varnothing$ initially
\State At each step add to $F$ a minimum-weight edge $e\not\in F$ such that $H$ remains a forest
\State Stop when $H$ is a spanning tree
\end{algorithmic}
\end{algorithm}

As far as we know, \cite{De Quincey-2019} first uses the Kruskal approach explicitly to get a $1/2$-approximation algorithm for MAX-CUT. The algorithm follows exactly the Kruskal procedure, but applies to unweighted graphs:

\begin{algorithm}[H]
\caption{De Quincey's Algorithm for MAX-CUT}\label{algo:De Quincey}
\begin{algorithmic}
\State \textbf{Input:} An unweighted graph $G=(V,E)$
\State \textbf{Output:} A partition $V=V_1\uplus V_2$
\State Keep a spanning forest $H=(V,F)$ of $G$, with $F=\varnothing$ initially
\State At each step add to $F$ an edge $e\not\in F$ randomly such that $H$ remains a forest
\State Stop when $H$ is a spanning tree
\State Return the partition $V=V_1\uplus V_2$ such that $F$ is within the cut $\delta(V_1)$
\end{algorithmic}
\end{algorithm}
Actually the corresponding relation tree here is just $H=(V,F)$ together with $s\equiv -1)$.

However, for MAX-CUT, we have better ways to avoid cycles: we could eliminate the correlation between the selected edges and the remaining ones by updating the graph. As long as we make the updating procedure properly, the maximum cut would be unchanged. Such an updating strategy has already been commonly used in solving MIN-CUT, and called ``Edge-Contraction~(EC)''~\cite{KS-1993}. Integrating the EC operation into the Kruskal algorithm, we would get another class of algorithms for MAX-CUT~\cite{Kahruman-2007}. Still we could have different criterions for the selection of the edge at each step, ``best-in'', ``worst-out'', and ``best-in-worst-out''. With the ``worst-out'' criterion, we immediately recover the original EC algorithm for MAX-CUT~\cite{Kahruman-2007}:

\begin{algorithm}[H]
\caption{EC Algorithm for MAX-CUT}\label{algo:EC}
\begin{algorithmic}
\State \textbf{Input:} A weighted graph $G=(V,E,w)$ with $w: E\to \mathbb{R}$
\State \textbf{Output:} A relation tree $H=(V,F,s)$
\State  $F\gets \varnothing$
\State $E'\gets E$
\While {$E'$ has more than one edges}
       \State Find the edge $(i,j)$ with the minimum weight in $E'$, add it to $F$
       \State $s_{ij}\gets +1$
       \State Shift all the edges $(i,k)$ with $k\ne j$ from $i$ to $j$
       \State Merge multiple edges in $E'$, and add up the weights
       \State $E'\gets E'\setminus \{ (i,j)\}$
\EndWhile
\State Add the last edge $e_0$ in $E'$  to $F$
\State $s_{e_0}\gets -1$
\State $H\gets (V,F,s)$
\end{algorithmic}
\end{algorithm}

Here we have rewritten the original EC algorithm in the language of relation tree on purpose. In this form the algorithm may be better called ``Edge Shifting'', rather than ``Edge Contraction'', but we keep the original name. When contracting the edges we actually have two choices, from $x$ to $y$ or from $y$ to $x$, but both give the same result.

Alternatively, one could use a ``best-in'' criterion for the selection of the edge at each step. The resulting algorithm is called ``Differencing Edge-Contraction''~(DEC)~\cite{Hassin-2021}, which we also rewritten in the language of relation tree here:

\begin{algorithm}[H]
\caption{DEC Algorithm for MAX-CUT}\label{algo:DEC}
\begin{algorithmic}
\State \textbf{Input:} A weighted graph $G=(V,E,w)$ with $w: E\to \mathbb{R}$
\State \textbf{Output:} A relation tree $H=(V,F,s)$
\State  $F\gets \varnothing$
\State $E'\gets E$
\While {$E'$ contains edges with positive weights}
       \State Find the edge $(i,j)$ with the maximum weight in $E'$, add it to $F$
       \State $s_{ij}\gets -1$
       \State Select one endpoint of $(i,j)$, say $i$
       \State Reverse the signs of the weights of all edges $(i,k)$ with $k\ne j$
       \State Shift all the edges $(i,k)$ with $k\ne j$ from $i$ to $j$
       \State Merge multiple edges in $E'$, and add up the weights
       \State $E'\gets E'\setminus \{(i,j)\}$
\EndWhile
\While {$E'$ is not empty}
       \State Randomly choose an edge $(i,j)$ in $E'$, add it to $F$
       \State $s_{ij}\gets +1$
       \State Shift all the edges $(i,k)$ with $k\ne j$ from $i$ to $j$
       \State Merge multiple edges in $E'$, and add up the weights
       \State $E'\gets E'\setminus \{(i,j)\}$
\EndWhile
\State $H\gets (V,F,s)$
\end{algorithmic}
\end{algorithm}

In contrast to EC, here the contraction direction does matter when the selected edges have positive weights, and affects the algorithm performance. A post-selection criterion is used in~\cite{Hassin-2021}, where the contraction direction is chosen to maximize the total weight of the resulting graph.

 Similar as SG3, we could take a delicate combination of the above two, and obtain a ``best-in-worst-out'' algorithm, which we call ``Signed-Edge-Contraction''~(SEC)~\cite{SAIII}:

\begin{algorithm}[H]
\caption{SEC Algorithm for MAX-CUT}\label{algo:SEC}
\begin{algorithmic}
\State \textbf{Input:} A weighted graph $G=(V,E,w)$ with $w: E\to \mathbb{R}$
\State \textbf{Output:} A relation tree $H=(V,F,s)$
\State  $F\gets \varnothing$
\State $E'\gets E$
\While {$E'$ is not empty}
       \State Find the edge $(i,j)$ such that the absolute value of its current weight $|\tilde w_{ij}|$ is maximized in $E'$
       \State add $(i,j)$ to $F$
       \State $s_{ij}\gets -\textrm{sgn}(\tilde w_{ij})$
       \For{$(i,k)\in E'$ with $k\ne j$}
           \State Multiply $\tilde w_{ik}$ by $s_{ij}$
           \State Shift $(i,k)$ from $i$ to $j$
           \State Merge multiple edges, and add up the weights
       \EndFor
       \State $E'\gets E'\setminus \{(i,j)\}$
\EndWhile
\State $H\gets (V,F,s)$
\end{algorithmic}
\end{algorithm}

One can see that SEC is extremely simply, compared to EC and DEC. The reason is that, both EC and DEC deal with different edge relations separately, while SEC deals with them in a uniform way. Moreover, SEC solves the contraction direction problem in DEC, since contraction along both directions gives weights with the same absolute value. This is why in the algorithm we do not have to make any choice with respect to the contraction direction. As an example, we apply SEC to the graph in Fig.~\ref{fig:G5}, and get the relation tree in Fig.~\ref{fig:G5-Tree}. An obvious fact is that, the tree in Fig.~\ref{fig:G5-Tree} contains more than one negative edges, while in both Fig.~\ref{fig:SG-Tree} and Fig.~\ref{fig:SG1-Tree} only one negative edge appears.

How about the time complexities of the Kruskal-class algorithms? Intuitively, since the original Kruskal algorithm has complexity $O(m\log m)$ ($m=|E|$), one may guess that all the algorithms here inherit such a complexity. However, while in the original Kruskal algorithm we have to make sure no cycle is formed at each step, this is not the case for the EC algorithms here. In spite of this, previous studies claim higher complexities: $O(mn^2)$ for De Quincey~\cite{De Quincey-2019}, $O(n^3)$ for EC~\cite{Kahruman-2007}, and $O(n^2\log n)$ for DEC~\cite{Hassin-2021}. In~\cite{SAIII} we claim that SEC has the complexity $O(n^2)$, the same as those SG algorithms. This is based on a ``label and update'' trick, which is commonly used in the Prim algorithm. We believe all the other algorithms in the Kruskal class could be improved to such a complexity.

\subsubsection{Approximation Ratio}

While in all Prim-class algorithms we could easily trace the change of the cut weight at each step, this is not the case in the Kruskal-class algorithms. Only when the selected edge is set as negative and included into the cut, this can be done by a careful analysis of all involved edges~\cite{De Quincey-2019,Hassin-2021}. In De Quincey's algorithm, all selected edges are included into the cut finally, and thus one could trace the change of the cut weight from beginning to end. With this, it is proved that De Quincey's Algorithm has an approximation ratio $1/2$~\cite{De Quincey-2019}. Obviously this method can not be generalized to the other three algorithms listed above, which makes the derivation of the approximation ratio complicated. The approximation ratio of EC is estimated to be larger than $1/3$~\cite{Kahruman-2007}, which is not so satisfactory. As for DEC, it is completely unknown~\cite{Hassin-2021}.

Quite unexpectedly, in~\cite{SAIII} it is proved that SEC has an approximation ratio $1/2$, just as the SG algorithms. The derivation is very straightforward and elegant, but to elucidate it clearly we will need a formulation of the problem more algebraic. This is done with the so-called ``Stabilizer formalism'' proposed by the quantum computing community. Using such an algebraic formulation one could easily obtain the cut weight at each step of the algorithm. In fact, the whole algorithm is first conceived in the stabilizer formalism, and then translated back into the graphic language. We will briefly introduce such a formalism in the next section.

\section{Stabilizer Formalism}

The stabilizer formalism is an important framework in quantum computation, first proposed in the late 90s~\cite{Gottesman-1996,Gottesman-1997,Gottesman-1998}. A nice introduction is given in~\cite{Nielsen-Chuang-2010}. Here we will only mention some structures and properties related to MAX-CUT.

\subsection{From Bit to Qubit}
As we show previously, using the spin variable the cut weight can be expressed as:
\begin{equation}
w(z)=\frac{1}{2}\sum_{i<j} w_{ij} (1-z_i z_j).
\end{equation}
Since the spin variable takes values in $\{+1,-1\}$, we could combine both values into a single diagonal matrix, the Pauli-$Z$ matrix:
\begin{equation}
Z\equiv\begin{pmatrix}
1&0\\
0&-1
\end{pmatrix}
\end{equation}
The corresponding eigenstates are defined as
\begin{equation}
|0\rangle\equiv \begin{pmatrix}
1\\
0
\end{pmatrix},\quad
|1\rangle \equiv \begin{pmatrix}
0\\
1
\end{pmatrix}
\end{equation}
so that
\begin{equation}
Z|0\rangle =|0\rangle,\quad Z|1\rangle =-|1\rangle,
\end{equation}
or more compactly,
\begin{equation}
Z|x\rangle =z |x\rangle, \quad z=1-2x.
\end{equation}

While $|x\rangle$ is simply the quantum promotion of the classical bit $x$, an intrinsic quantum bit $|\psi\rangle$ could be any complex linear combination of the classical state $|x\rangle$:
\begin{equation}
|\psi\rangle\equiv \alpha|0\rangle+\beta|1\rangle,
\end{equation}
satisfying the normalization condition:
\begin{equation}
\langle\psi|\psi\rangle\equiv|\alpha|^2+|\beta|^2=1.
\end{equation}
So a single qubit is a normalized vector in $\mathbb{C}^2$. If we further ignore an overall phase factor of $|\psi\rangle$, the single-qubit state could be parameterized as
\begin{equation}
|\psi\rangle\simeq \cos \frac{\theta}{2}|0\rangle+\mathe^{i\varphi}\sin \frac{\theta}{2}|1\rangle.
\end{equation}
This is the Bloch sphere representation of a single qubit, with $(\theta, \phi)$ the elevation and azimuth. In such a representation the classical bit state $|x\rangle$ lies on the $z$-axis. The states lying on the $x$ and $y$-axis are eigenstates of the Pauli-$X$ and Pauli-$Y$ matrices:
\begin{equation}
X\equiv\begin{pmatrix}
0&1\\
1&0
\end{pmatrix},\quad
Y\equiv\begin{pmatrix}
0&-i\\
i&0
\end{pmatrix}
\end{equation}
To transform states from one axis to another, we will need another two matrices:
\begin{equation}
H\equiv \frac{1}{\sqrt{2}}\begin{pmatrix}
1&1\\
1&-1
\end{pmatrix},\quad
S\equiv\begin{pmatrix}
1&0\\
0&i
\end{pmatrix},
\end{equation}
called Hadamard matrix and phase matrix.

When we have $n$ qubits, the entire state is described by a normalized vector in the tensor product space ${(\mathbb{C}^{2}})^{\otimes n}$. We will omit the tensor product symbol between different qubits later, both for the states and the matrices.

\subsection{Stabilizer States}
While MAX-CUT is indeed a classical problem and the solution is definitely a classical $n$-bit state $x$, it turns out to be more convenient to consider more general $n$-qubit states. As a hint of this, $x$ and its negation $\neg x$, both describing the same cut, could be simultaneously represented with a single $n$-qubit state.

But we do not really want to consider all $n$-qubit states, we just need some specific states of them. To specify such a subspace, we first define the Pauli group. The one qubit Pauli group is defined to be:
\begin{equation}
G_1\equiv \{ \pm I, \pm iI, \pm X,\pm iX, \pm Y, \pm iY, \pm Z, \pm iZ\}.
\end{equation}
Then the $n$-qubit Pauli group is defined as its $n$-fold tensor product:
\begin{equation}
G_n\equiv G_1^{\otimes n}
\end{equation}
So an element in $G_n$ is actually an $n$-fold tensor product of Pauli matrices, together with an overall factor of $\pm 1,\pm i$.

Next we seek a special class of subgroups $S_n$ of $G_n$ to characterize the special states we want. These are defined as:
\\
\textbf{Definition 2}:\\
A \textbf{stabilizer group} $S_n$ is a subgroup of $G_n$ such that:\\
1. $S_n$ is commutative;\\
2. $-I\not\in S_n$;\\
3. $S_n$ has exactly $n$ generators, or $S_n=<s_1,...,s_n>$.\\
Correspondingly, we define the stabilizer state as:
\\
\textbf{Definition 3}:\\
A $n$-qubit state $|\psi\rangle$ is called a \textbf{stabilizer state} if there is a stabilizer group $S_n$, such that:
\begin{equation}
s|\psi\rangle=|\psi\rangle,\quad \forall s\in S_n.
\end{equation}
Let us give some typical examples of this. The well-known Einstein-Podolsky-Rosen~(EPR) pair/Bell state
\begin{equation}
|\psi\rangle=\frac{|00\rangle+|11\rangle}{\sqrt{2}}
\end{equation}
is a stabilizer state with the stabilizer group $\langle ZZ,XX\rangle$. Also any classical state $|x\rangle$ is a stabilizer state, with individual stabilizer generator given by $Z$ or $-Z$.

\subsection{SEC revisited}
Now we try to formulate MAX-CUT in the stabilizer formalism. First, we could promote the weight function $w(z)$ into a matrix function, or in physical language, the Hamiltonian operator:
\begin{equation}
H(Z)=-w(Z)=-\frac{1}{2}\sum_{i<j} w_{ij}(1-Z_iZ_j).\label{eq.wZ}
\end{equation}
Here an additional minus has been included so that the maximum cut corresponds to the ground state of $H(Z)$ with the lowest eigenvalue. Since MAX-CUT is a classical problem, the corresponding ground state must be a classical state, and thus a stabilizer state. Therefore we could approach the ground state within the subspace of stabilizer states.

Such a state could be engineered by choosing each stabilizer generators one by one. The individual terms in the Hamiltonian would be natural candidates for the stabilizers. The criterion for the choice of the generators is such that the Hamiltonian is as negative as possible. If we do this greedily, we immediately recover the SEC algorithm. To see this, let's define the set of nontrivial Pauli operators appearing in the Hamiltonian as $P$. Initially, we have
\begin{equation}
P=\{Z_iZ_j|(i,j)\in E\}.
\end{equation}
We also define the coefficients of those Pauli terms as a function on $P$:
\begin{equation}
c_P: P\to \mathbb{R},
\end{equation}
initiated as $c_{ij}=w_{ij}/2$. Finally, we use $H_0$ to store the constant term of $H(Z)$, which is initially $H_0=-\frac{1}{2}\sum_{i<j} w_{ij}$. Each time a new Pauli term (or its opposite) is chosen as a generator, it could be identified as the identity operator, and $H_0$ changes accordingly. With these definitions, we could reformulate SEC as:

\begin{algorithm}[H]
\caption{SEC Algorithm for MAX-CUT}\label{algo:SEC'}
\begin{algorithmic}
\State \textbf{Input:} A weighted graph $G=(V,E,w)$ with $w: E\to \mathbb{R}$
\State \textbf{Output:} Generators for a stabilizer group $S_n$ and the cut weight $w(S)$
\State Initiate the Pauli set as $P\gets \{Z_iZ_j|(i,j)\in E\}$
\State Initiate the coefficient set as $c_P\gets \{c_{ij}=w_{ij}/2|(i,j)\in E\}$
\State $H_0\gets -\frac{1}{2}\sum_{i<j} w_{ij}$
\For {$l=1:n-1$}
       \State Find $Z_iZ_j$ in $P$ such that the absolute value of its current coefficient $|\tilde c_{ij}|$ is maximized in $c_P$
       \State $s_l\gets -\textrm{sgn}(\tilde c_{ij})Z_iZ_j$
       \For{$k\ne j$ such that $Z_iZ_k \in P$}
            \State Multiply $\tilde c_{ik}$ by $-\textrm{sgn}(\tilde c_{ij})$
            \State Multiply $Z_iZ_k\in P$ by $Z_iZ_j$
            \State Merge identical terms in $P$ and add up the coefficients in $c_P$
       \EndFor
       \State $P\gets P\setminus \{Z_iZ_j\}$
       \State $H_0\gets H_0-|\tilde c_{ij}|$
\EndFor
\State $s_n\gets Z_1$
\State $S_n\gets <s_1,s_2,...,s_n>$
\State $w(S)\gets -H_0$
\end{algorithmic}
\end{algorithm}

Almost all the steps here parallel those in SEC. We could even prove that at each step, the current weights of edges and current coefficients of Pauli terms are related to each other exactly:
\begin{equation}
\tilde c_{ij}=\tilde w_{ij}/2.
\end{equation}
Moreover, each edge $(i,j)$ in the relation tree, together with its sign assignment $s_{ij}$, corresponds to a stabilizer generator $s_{ij}Z_iZ_j$ here. This gives in total $n-1$ stabilizer generators. Of course these are enough to determine the cut. We formally include an extra generator $Z_1$ to fix the stabilizer group completely. To explicitly give the cut, one could simply replace all $Z_i$ matrices in these generators by its eigenvalue $z_i$, and solve the $n$ coupled equations $s_l=1,~(l=1,...n)$ accordingly. Of course this is equivalent to scanning the corresponding relation tree.

However, there is an additional operation in the above algorithm: we could easily trace the change of the cut at each step, which allows us to easily derive its approximation ratio. As we said, this is not easy for all the Kruskal-class algorithms.

\subsubsection{Approximation Ratio}

From the above algorithm, we could actually get a compact expression for the final cut. For this, let us denote the current coefficient selected at step $l$ as $\tilde c^l$. Then
\begin{equation}
w(S)=-H_0=\frac{1}{2}\sum_{i<j} w_{ij}+\sum_{l=1}^{n-1} |\tilde c^{l}|.
\end{equation}
Alternatively, since the $n-1$ generators and edges in the relation tree $H=(V,F,s)$ are in one-to-one correspondence, we could rewrite the weight as
\begin{equation}
w(F)=\frac{1}{2}\sum_{i<j} w_{ij}+\frac{1}{2}\sum_{(i,j)\in F}|\tilde w_{ij}|.
\end{equation}
Again $\tilde w_{ij}$ is the current weight of edge $(i,j)$ when it is chosen. From this expression one immediately gets the conclusion:
SEC has approximation ratio $1/2$.

How about the other Kruskal-class algorithms? In analyzing the SG algorithms we know that all of them have approximation ratio $1/2$, as long as we assign each vertex greedily. Different strategies to select the vertices do not affect the approximation ratio theoretically. Could we get the same conclusion here? The answer is yes. To ensure a $1/2$ approximation ratio here, we
must make sure that at each step, the cut weight is changed by a positive quantity, rather than a negative one. This can be guaranteed if we properly assign the sign $s_{ij}$ of the edge $(i,j)$: it must be opposite to the sign of the current weight $\tilde w_{ij}$. The meaning of this rule is clear: if the edge has a positive weight, we include it into the cut; otherwise we discard it. As long as this rule is obeyed, the cut weight will always be of the same form as in the above expression. Different strategies to select the edges only change the relation tree itself, or the summation indices, not the form of the above expression.

Now we could check if such an edge-assignment rule is obeyed in all the Kruskal-class algorithms. Clearly, De Quincey, DEC and SEC all obey this rule, and all have the approximation ratio $1/2$. EC could violate this rule severely, and thus may not guarantee an approximation ratio of $1/2$.

\subsection{SG revisited}\label{ss:SG}
We could also rewrite the SG algorithms in such a formalism, which also helps to determine the approximation ratio. Let's take the original SG algorithm as an example.

Assume that we have ordered the vertices as $1,2,...,n$. For each vertex $1\le j \le n$, we have to choose the stabilizer as $Z_j$ or $-Z_j$. Correspondingly, the eigenvalue of the $Z_j$ matrix would be $z_j=1$ and $z_j=-1$. To make the criterion clear, we rewrite eq.(~\ref{eq.wZ}) as
\begin{eqnarray}
w(Z)&=&\frac{1}{2}\sum_{i<j} w_{ij}(1-Z_iZ_j)\nonumber\\
    &=& \frac{1}{2}\sum_{i<j} w_{ij}- \frac{1}{2}\sum^n_{j=2} (\sum_{i<j} w_{ij}Z_i)Z_j
\end{eqnarray}
At step $j$, all the eigenvalues $z_i,i<j$ would have been determined. So we could formally write the cut weight as
\begin{equation}
w(Z)= \frac{1}{2}\sum_{i<j} w_{ij}- \frac{1}{2}\sum_{j=2}^n (\sum_{i<j} w_{ij}z_i)Z_j
\end{equation}
Since $Z_1$ now does not appear, we may fix $z_1$ arbitrarily. At step $j>1$, we make the choice as
\begin{equation}
z_j=-\textrm{sgn}(\sum_{i<j} w_{ij}z_i),\quad 2\le j \le n,
\end{equation}
It is easy to check that such a choice agrees with that in the original SG algorithm when $w_{12}\ge0$..
Then we get the final cut weight:
\begin{equation}
w(z)= \frac{1}{2}\sum_{i<j} w_{ij}+\frac{1}{2}\sum_{j=2}^n \big|\sum_{i<j} w_{ij}z_i\big|.
\end{equation}
Obviously the weight is larger than one half of the total weight, and thus the approximation ratio is $1/2$. Notice that the above procedure is slightly different from the original SG algorithm: it starts with a single vertex, rather than two vertices or an edge. This seems to be more natural, and could properly deal with the situation when $w_{12}<0$. We will come back to this later.

\subsection{\textit{ADAPT-Clifford}}
Now we turn to a recently proposed algorithm for MAX-CUT, \textit{ADAPT-Clifford}~\cite{ADAPT-Clifford}, which is also formulated in the stabilizer formalism. The algorithm construct the stabilizer state explicitly, and thus is a little complicated. Let us introduce the algorithm first, and then explain the steps in graphical language.

\begin{algorithm}[H]
\caption{\textit{ADAPT-Clifford} Algorithm for MAX-CUT}\label{algo:ADAPT-Clifford}
\begin{algorithmic}
\State \textbf{Input:} A weighted graph $G=(V,E,w)$ with $w: E\to \mathbb{R}$
\State \textbf{Output:} A stabilizer state $\psi_n$ and the cut $V=A\uplus \bar A$
\State Prepare the initial state $|\psi_0\rangle=H^n|0^n\rangle$
\State Select a vertex $i_1$ and prepare the state $|\psi_1\rangle=Z_{i_1}|\psi_0\rangle$
\State $V'\gets V\setminus \{i_1\}$
\State Find the edge $(i_1,i_2)$ of largest positive weight incident with $i_1$
\State $|\psi_2\rangle\gets \mathe ^{i \frac{\pi}{4}Z_{i_2}Y_{i_1}}|\psi_1\rangle$
\State $V'\gets V'\setminus \{i_2\}$
\For {$r=3:n$}
     \For{$i\in V'$}
               \State $g_{i i_1}\gets -\sum_{l\in V\setminus V'} w_{il}\langle Z_lX_iZ_{i_1}\rangle_{\psi_{r-1}}$
               \State $g_{i i_2}\gets -\sum_{l\in V\setminus V'} w_{il}\langle Z_lX_iZ_{i_2}\rangle_{\psi_{r-1}}$
               \State $\textrm{score}(i) \gets \max\{g_{i i_1}, g_{i i_2}\}$
     \EndFor
     \State $i^*\gets \arg \max_{i\in V'}{\textrm{score}(i)}$
     \If{$g_{i^* i_1}>0$}
              \State $|\psi_r\rangle\gets \mathe ^{i \frac{\pi}{4}Y_{i^*}Z_{i_1}}|\psi_{r-1}\rangle$
     \Else    \State $|\psi_r\rangle\gets \mathe ^{i \frac{\pi}{4}Y_{i^*}Z_{i_2}}|\psi_{r-1}\rangle$
     \EndIf
     \State $V'\gets V'\setminus \{i^*\}$
\EndFor
\State Measure $|\psi_n\rangle$ to get the classical state $|x\rangle$
\State $A\gets \{i\in V|x_i=1\}$
\State $\bar A\gets \{i\in V|x_i=0\}$
\end{algorithmic}
\end{algorithm}

To see how the cut is actually constructed, it would be better to trace the stabilizer generators step by step. The update of the generators has been partially explained in~\cite{ADAPT-Clifford}, which we will elucidate in more detail.

Step 0:\\
Initially we have
\begin{equation}
|\psi_0\rangle=H^n|0^n\rangle=|+^n\rangle,
\end{equation}
where $+\rangle\equiv \frac{1}{\sqrt{2}}(|0\rangle+|1\rangle$. Since $X|+\rangle=|+\rangle$, the initial generators are $X_1,X_2,...,X_n$. That is to say, all the vertices are undetermined.

Step $1$:\\
Then we randomly select an initial vertex $i_1$, and apply $Z_{i_1}$ to $|\psi_0\rangle$. This turns generator $X_{i_1}$ into $-X_{i_1}$, while the other generators are unchanged. We exclude $i_1$ from $V$ to indicate vertex $i_1$ has been dealt with, and denote the remaining set as $V'$.

Step $2$:\\
Choose vertex $i_2$ such that the weight of edge $(i_1,i_2)$ is maximized among all the incident edges of $i_1$. In special cases we may still have $w_{i_1 i_2}<0$, but we may neglect such subtlety and assume $w_{i_1 i_2}>0$. We then apply $\mathe ^{i \frac{\pi}{4}Z_{i_2}Y_{i_1}}$ to the previous state and get
\begin{equation}
|\psi_2\rangle= \mathe ^{i \frac{\pi}{4}Z_{i_2}Y_{i_1}}|\psi_1\rangle.
\end{equation}
A little work shows that with this operation, generator $-X_{i_1}$ changes into $-X_{i_1}X_{i_2}$, and $X_{i_2}$ changes into $-Z_{i_1}Z_{i_2}$. The other generators remain the same. This means vertices $i_1$ and $i_2$ are to be put into different sets. Following SG3, we still use $V_1$ and $V_2$ to denote the two sets, and initiate $V_1=\{i_1\}$ and $V_2=\{i_2\}$. Then the remaining vertex set is $V'=V\setminus (V_1\uplus V_2)$. We may further write the generator $-X_{i_1}X_{i_2}$ as $-\prod_{l\in V_1\uplus V_2} X_l$.

Step $3$ to $n$:\\
Now at each step we have to determine which vertex to select. The criterion to make the choice is based on the score of each undetermined vertex $i$ in $V'$:
\begin{eqnarray}
&&g_{i i_1}= -\sum_{l\in V \setminus V'} w_{il}\langle Z_lX_iZ_{i_1}\rangle_{\psi_{r-1}}\nonumber\\
&&g_{i i_2}= -\sum_{l\in V \setminus V'} w_{il}\langle Z_lX_iZ_{i_2}\rangle_{\psi_{r-1}}\nonumber\\
&&\textrm{score}(i) =  \max\{g_{i i_1}, g_{i i_2}\}.\label{eq.ADAPT}
\end{eqnarray}
We need to simplify the above expressions to see what do they really calculate. Firstly, since $i\in V'$ has stabilizer $X_i$, so $X_i$ can be trivially factored out. Next, since $V\setminus V'=V_1\uplus V_2$, we may do the calculation separately for $V_1$ and $V_2$, and find
\begin{eqnarray}
&&\langle Z_lZ_{i_1}\rangle_{\psi_{r-1}}=1, ~~~~\langle Z_lZ_{i_2}\rangle_{\psi_{r-1}}=-1,\quad l\in V_1,\nonumber\\
&&\langle Z_lZ_{i_1}\rangle_{\psi_{r-1}}=-1, ~~\langle Z_lZ_{i_2}\rangle_{\psi_{r-1}}=1,~~\quad l\in V_2
\end{eqnarray}
With these relations we may simplify the expressions in Eqs.~\ref{eq.ADAPT} as:
\begin{eqnarray}
&&g_{i i_1}= w(i,V_2)-w(i,V_1)\nonumber\\
&&g_{i i_2}=w(i,V_1)-w(i,V_2)\nonumber\\
&&\textrm{score}(i) =\max\{g_{i i_1}, g_{i i_2}\}=|w(i,V_1)-w(i,V_2)|.
\end{eqnarray}
So the score function is exactly the same as SG3! This guarantees that at each step the same vertex $i^*$ is chosen. Then we have to check if the same assignment as SG3 has been performed. Without loss of generality, assume $g_{i^* i_1}>0$ or $w(i^*,V_2)>w(i^*,V_1)$, so we apply the matrix $\mathe ^{i \frac{\pi}{4}Y_{i^*}Z_{i_1}}$ to the previous state. Some matrix multiplication shows that it turns generator $X_{i^*}$ into $Z_{i^*}Z_{i_1}$. This means we have included $i^*$ into $V_1$, exactly the same as in SG3. The operation also updates another generator $-\prod_{l\in V_1\uplus V_2} X_l$, which does not affect the cut.

Finally, the stabilizer state $|\psi_n\rangle$ is determined by the generators
\begin{equation}
-Z_{i_1}Z_{i_2}, Z_rZ_{i_1}/Z_rZ_{i_2}~(r=3,...,n), -\prod_{l\in V} X_l.
\end{equation}
Due to the presence of the generator $-\prod_{l\in V} X_l$, $|\psi_n\rangle$ is a superposition of a state $|x\rangle$ and its negation $|\neg x\rangle$:
\begin{equation}
|\psi_n\rangle = \frac{1}{\sqrt{2}}(|x\rangle-|\neg x\rangle).
\end{equation}
Measuring $|\psi_n\rangle$ gives either $x$ or $\neg x$, and correspondingly $A$. Since the final cut $V=A\uplus \bar A$ must be the same as $V=V_1 \uplus V_2$, we would simply get $A=V_1$ or $A=V_2$.

In conclusion, the \textit{ADAPT-Clifford} algorithm is just a reformulation of SG3 in the stabilizer formalism. However, \cite{ADAPT-Clifford} reports a rather high complexity, $O(n^5)$, while SG3's complexity is of $O(n^2)$. This means, most of operations in~\cite{ADAPT-Clifford} are redundant, and could be removed or simplified. Nevertheless, the \textit{ADAPT-Clifford} algorithm does give some hints that all the SG algorithms, as of Prim-class, should better start from a vertex rather than an edge, even though this may introduce some randomness.

\section{Numerical Performance}

Now we turn to the numerical performance of these algorithms. Previous study in~\cite{Kahruman-2007} shows that for the SG algorithms, the ``best-in-worst-out'' version, namely SG3, performs much better than the others. Similarly, SEC as a ``best-in-worst-out'' synthesis of EC and DEC, is expected to perform best in the Kruskal class. So we will mainly focus on these two algorithms. But before doing the numerical calculation, we would like to modify the initialization of SG3 slightly, following the choice made in~\cite{ADAPT-Clifford}. Namely, we would start from an initial vertex, as the original Prim algorithm does, rather than from the maximum-weight edge. Then the algorithm reads:

\begin{algorithm}[H]
\caption{SG3 Algorithm for MAX-CUT}\label{algo:SG3}
\begin{algorithmic}
\State \textbf{Input:} A weighted graph $G=(V,E,w)$ with $w: E\to \mathbb{R}$
\State \textbf{Output:} A partition $V=V_1\uplus V_2$ and the cut weight $w(V_1,V_2)$
\State Choose an initial vertex $r$
\State $V_1\gets \{r\}, V_2\gets \varnothing$
\State $w(V_1,V_2) \gets 0$
\State $V'\gets V\setminus \{r\}$
\For {$j=1:n-1$}
     \For {$i\in V'$}
          \State $w(i,V_1)=\sum_{(i,j)\in E,j\in V_1}w(i,j)$
          \State $w(i,V_2)=\sum_{(i,j)\in E,j\in V_2}w(i,j)$
          \State $\textrm{score}(i)= |w(i,V_1)-w(i,V_2)|$
     \EndFor
     \State
     $i^*\gets \arg \max_{i\in V'} {\textrm{score}(i)}$
         \If {$w(i^*,V_1)> w(i^*,V_2)$}
              \State $V_2\gets V_2\cup \{i^*\}$
         \Else
              \State $V_1\gets V_1\cup \{i^*\}$
         \EndIf
     \State $V'\gets V'\setminus \{i^*\}$
     \State $w(V_1,V_2)\gets w(V_1,V_2)+\max\{w(i^*,V_1),w(i^*,V_2)\}$
\EndFor
\end{algorithmic}
\end{algorithm}

Now the performance of the algorithm would depend on the choice of the initial vertex $r$. We could use an exhaustive search to find the best choice for $r$, at the price of increasing the time complexity by one factor of $n$. We call such an approach ``deterministic SG3''~(SG3-d). Alternatively, we may choose $r$ randomly, and then repeat the whole algorithm $t$ times. We call this ``randomized SG3''~(SG3-r). we fix $t$ to be $t=[2\log_2 n]$, so that it could be much faster than the deterministic one, while somehow maintaining the performance. Since the modified SG3 now differs from \textit{ADAPT-Clifford} only by complexity, some of the following results for SG3-d and SG3-r in the small-size region may coincide with those in~\cite{ADAPT-Clifford}.

\subsection{Complete graphs}

\subsubsection{Positive weights}

\begin{figure}[H]
\centering
	\includegraphics[width=\textwidth]{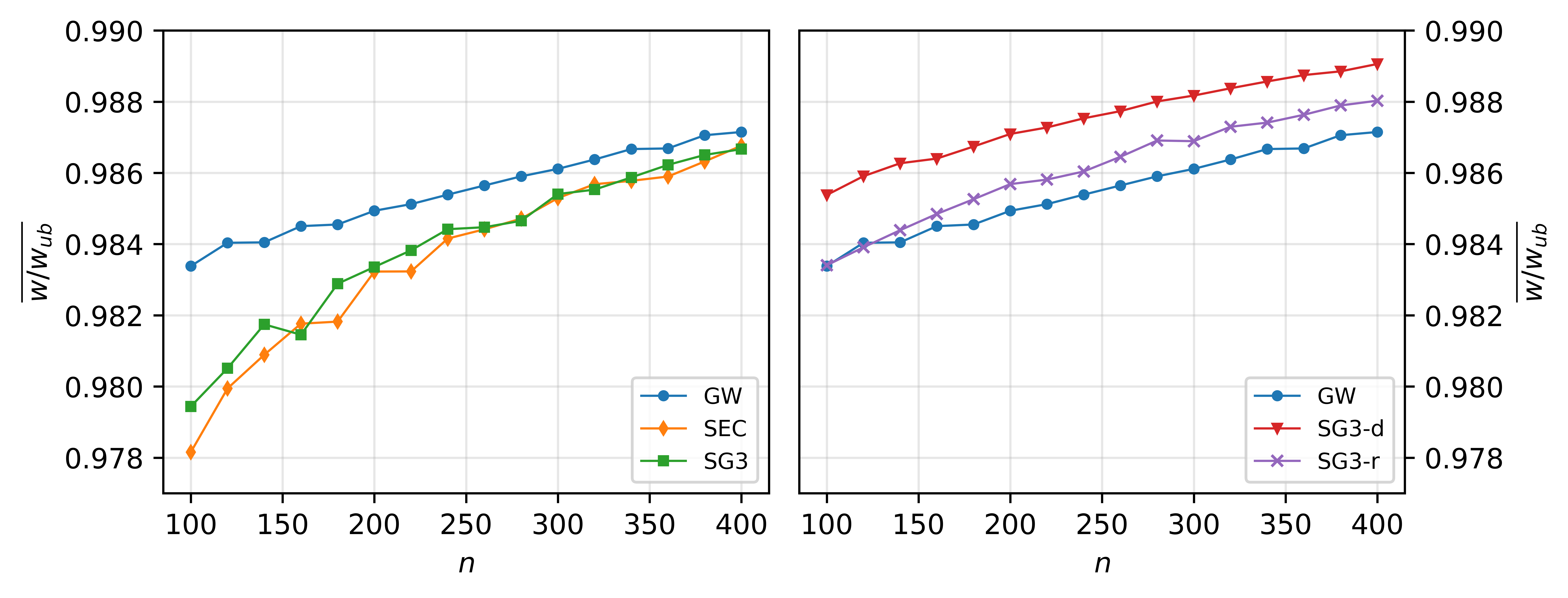}
\caption{\it Average performance of various algorithms for complete graphs with positive weights. For each $n$, we generate $50$ instances with the weights sampled independently from $U[0,1]$, calculate the ratio $w/w_{ub}$ for each instance, and finally take the average over all instances. Here $w_{ub}$ is the SDP upper bound from the GW procedure. Left: results from the original SG3 and SEC algorithm, in comparison with GW. Right: comparison of results from the improved SG3-d and SG3-r algorithms and GW.}\label{fig:Complete-Graph}
\end{figure}

We start with complete graphs with positive weights, and sample each weight independently from the uniform distribution $U[0,1]$. The graph size $n$ is chosen to range from $100$ to $400$. For comparison, we first run the SDP-based GW algorithm using the Julia implementation~\cite{GW-Julia}, fixing the number of rounding times of GW to be $100$~\footnote{The performance of GW could be slightly improved if we increase the rounding times to $10^3$ or $10^4$~\cite{ADAPT-Clifford}. However, since here we only use GW for graphs with hundreds of vertices, repeating the rounding procedure $100$ times seems reasonable.}. For each $n$, we generate $50$ instances, and run GW on each of them. The ratio of the predicted cut weight to the SDP upper bound, $w_{ub}$, is used to characterize the performance on each instance. This ratio is then averaged over all $50$ instances to give the final prediction. All the other algorithms are analyzed in the same way, taking the corresponding SDP upper bound $w_{ub}$ as reference. The final results for all the algorithms are presented in Fig.~\ref{fig:Complete-Graph}.

From the left figure, we see that the results from the original SG3 algorithm and SEC are almost the same. They are not as good as GW, but as the graph size increases they gradually get close. When $n$ is bigger than $300$, they are almost as accurate as GW. Comparing the curves from the left and right figures, one finds significant improvement from SG3 to SG3-r and SG3-d. Both of them outperform GW in the whole size region, and SG3-d performs extremely well. At $n=400$, the prediction of SG3-d is almost $99\%$ of the SDP upper bound, indicating that it could be very close to the optimal value. These results demonstrate that, for the vertex-oriented algorithm such as SG3, starting from a maximum-weight edge is indeed not a good choice. Also, the randomized algorithm SG3-r could be further improved, for example by increasing the number of repeating times from $[2\log_2 n]$ to $[\sqrt{n}]$.

In short, modifying the initialization from an edge to a vertex greatly improves the performance of the SG3 algorithm, making it much better than GW and SEC. The superiority of SG3-d/r over SEC on complete graphs is expected: Prim-class algorithms ought to perform better than Kruskal-class algorithms for dense graphs. Still, their superiority over GW is unexpected and noteworthy.

\subsubsection{Signed weights: the Sherrington-Kirkpatrick model}

Next we consider graphs with both positive and negative weights. We will focus on a specific model, the Sherrington-Kirkpatrick~(SK) model, described by the following Hamiltonian on a complete graph:
\begin{equation}
H_{\rm{SK}}=\sum_{i<j} w_{ij} z_i z_j,
\end{equation}
where each weight is sampled independently from a symmetric distribution satisfying
\begin{equation}
\overline{ w_{ij}}=0,\quad \overline {w_{ij}^2}=1, \quad w_{ij}=w_{ji}.
\end{equation}
As a special instance of the Ising-type Hamiltonian, the SK model could be related to a MAX-CUT problem for a graph with edge weights $\{w_{ij}\}$. One could first solve the MAX-CUT problem and then extract the ground state energy $E_{\rm{SK}}$ through Eq.~(\ref{eq:wz}). Alternatively, one could exactly solve the model in the thermodynamic limit~\cite{Parisi-1979}, getting
\begin{equation}
\lim_{n\to \infty} \frac{E_{\rm{SK}}}{n^{3/2}}=\Pi^*\approx-0.763.
\end{equation}
This limiting value is called the Parisi value. Some other approaches give rise to less accurate limiting values. For example, using similar derivation as in subsection (\ref{ss:SG}) one finds for the original SG algorithm~\cite{Aizenman-1987,Rigetti-2023}:
\begin{equation}
\Pi_{\rm{SG}}=-\frac{2}{3}\sqrt{\frac{2}{\pi}}\approx -0.532.
\end{equation}
The SDP method gives a better value~\cite{Aizenman-1987}:
\begin{equation}
\Pi_{\rm{SDP}}=-\frac{2}{\pi}\approx -0.637.
\end{equation}
We will use all three limiting values as references to assess various algorithms.

\begin{figure}[H]
\centering
	\includegraphics[width=0.8\textwidth]{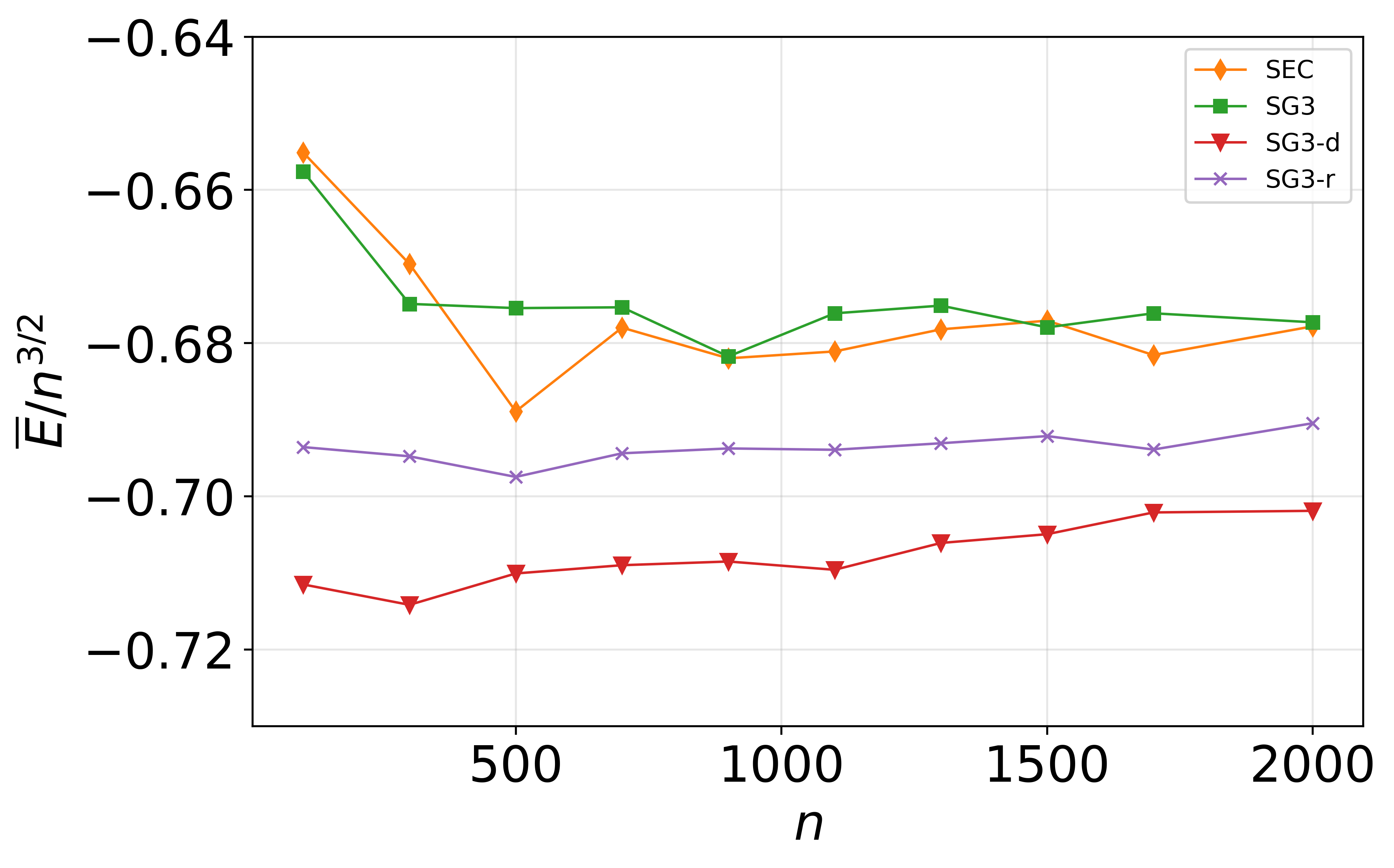}
\caption{\it Average performance of various algorithms for the SK model. For each $n$, we generate $10$ instances with $\{w_{ij}\}$ sampled independently from $\mathcal{N}(0,1)$, calculate the regularized energy $\bar{E}/n^{3/2}$ for each instance, and finally take the average over all instances. }\label{fig:SK-2000}
\end{figure}

\begin{figure}[H]
\centering
	\includegraphics[width=0.8\textwidth]{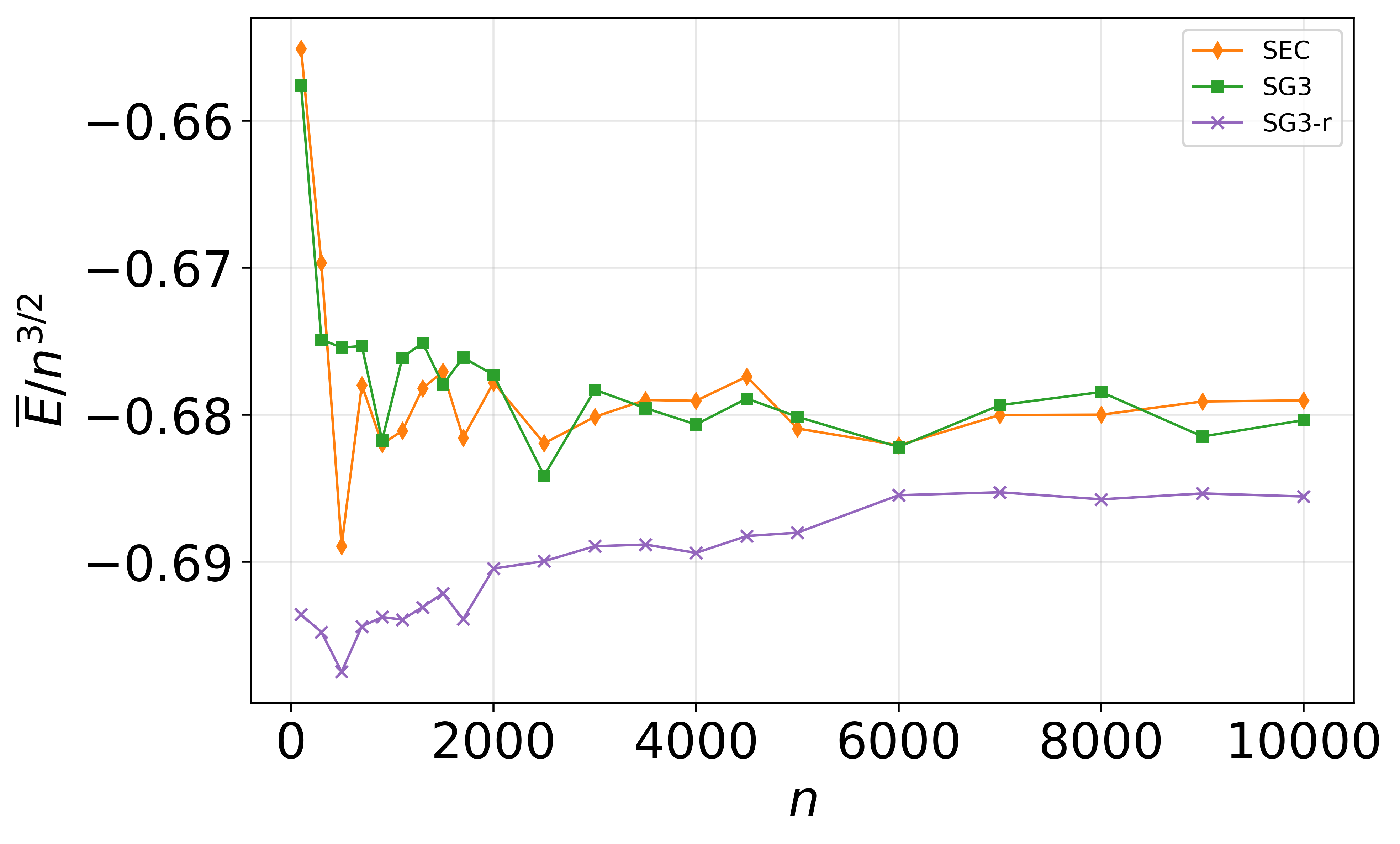}
\caption{\it Average performance of various algorithms for the SK model. For each $n$, we generate $10$ instances with $\{w_{ij}\}$ sampled independently from $\mathcal{N}(0,1)$, calculate the regularized energy $\bar{E}/n^{3/2}$ for each instance, and finally take the average over all instances. }\label{fig:SK-10000}
\end{figure}

We generate $10$ instances for each graph size $n$, and sample the edge weights independently from the normal distribution $\mathcal{N}(0,1)$. We calculate the average cut value over all instances, and then extract the ground state energy of the SK model through the relation (\ref{eq.wz}). For SEC, SG3 and SG3-r, we do the calculation for $n$ as large as $10000$, while for SG3-d we manage to perform the calculation up to $n=2000$. The results are summarized in Fig.~\ref{fig:SK-2000} and Fig.~\ref{fig:SK-10000}.


From the above figures we can see similar patterns as the previous case with only positive weights: the original SG3 and SEC have nearly the same performance, while SG3-d outperform them a lot. SG3-r stays somewhere in between, and may be further improved by increasing the repeating times. The limiting value of SG3 and SEC is about $-0.68$, already surpasses the SDP limit. SG3-d seems to converge at a value around $-0.70$, while SG3-r manages to get a value around $-0.686$. Both of them are still far from the Parisi value, making clear that more advanced techniques are further needed to approach the true ground state. We also notice that the results at small $n$ suffer from large finite-size fluctuations, and naively extrapolating those results to large $n$ could lead to inaccurate answers~\cite{ADAPT-Clifford}.


\subsection{Sparse graphs}
While the vertex-oriented Prim-class algorithms may perform better for dense graphs, one expects the edge-oriented Kruskal-class to win out for sparse enough graphs. In particular, when the edge number $m$ is of the same order as the vertex number $n$, the Kruskal-class algorithm should exhibit some advantage. Nevertheless, the dependence of different algorithms on the sparsity could be entangled with the dependence on the weight distribution. Therefore, we consider both weighted and unweighted graphs next, all with fixed average degree.

\subsubsection{$k$-regular graphs}

We start with $k$-regular random graphs. When $k$ is large enough, we should get similar results as the previous case of complete graphs. To confirm this, we consider random $8$-regular graphs first. We generate $40$ instances for various $n$ in the range $[60,410]$, calculate the ratio of the obtained cut to the total weight for each instance, and finally take the average over all instances. Here we choose the total weight rather than the SDP upper bound to calculate the cut ratio, because this ratio could be estimated with a large-$k$ expansion, and stays almost as a constant when $n$ is varied~\cite{Dembo-2017}.

\begin{figure}[H]
\centering
	\includegraphics[width=0.85\textwidth]{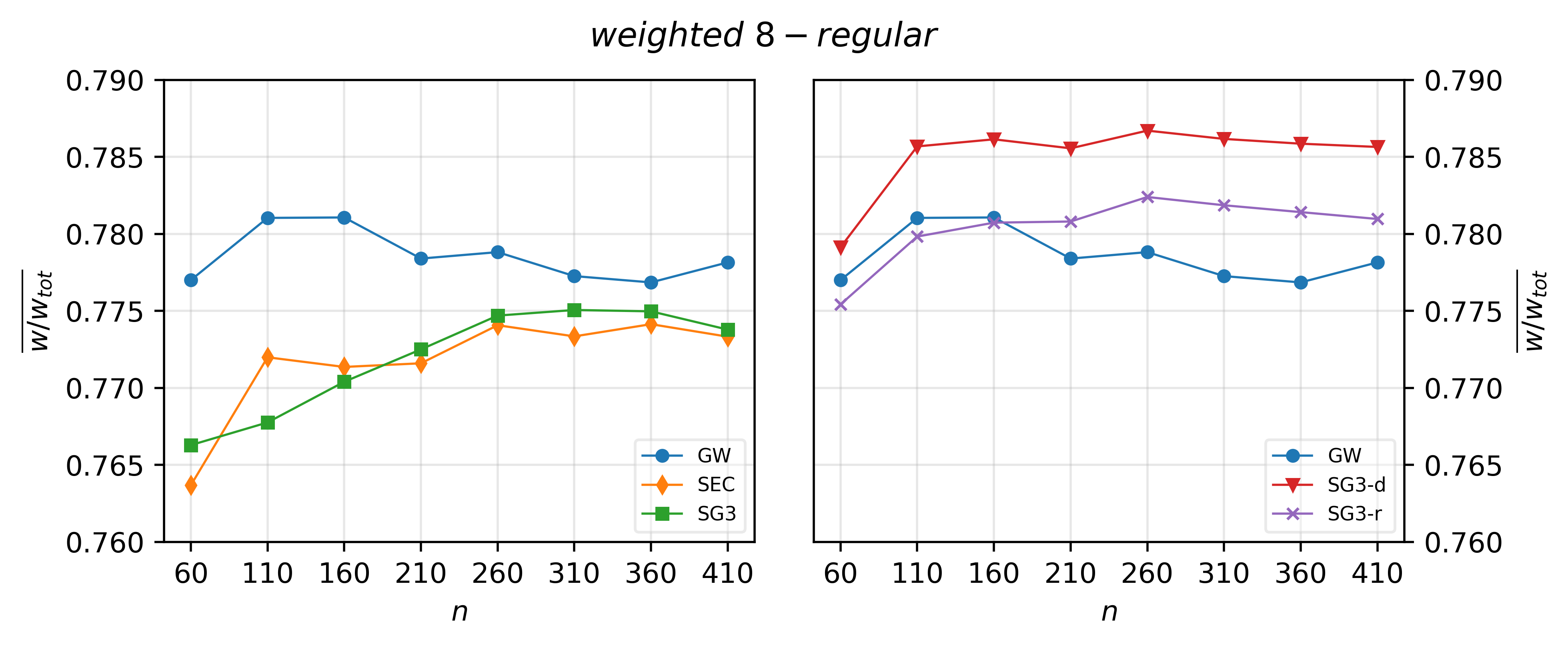}
\caption{\it Average performance of various algorithms for the weighted $8$-regular graphs. For each $n$, we generate $40$ instances with $\{w_{ij}\}$ sampled independently from $U[0,1]$, calculate the cut ratio $w/w_{tot}$ for each instance, and finally take the average over all instances. Left: results from SEC, SG3 and GW are plotted; Right: results from SG3-d and SG3-r and GW results are shown.}\label{fig:wt-8r}
\end{figure}

\begin{figure}[H]
\center
	\includegraphics[width=0.85\textwidth]{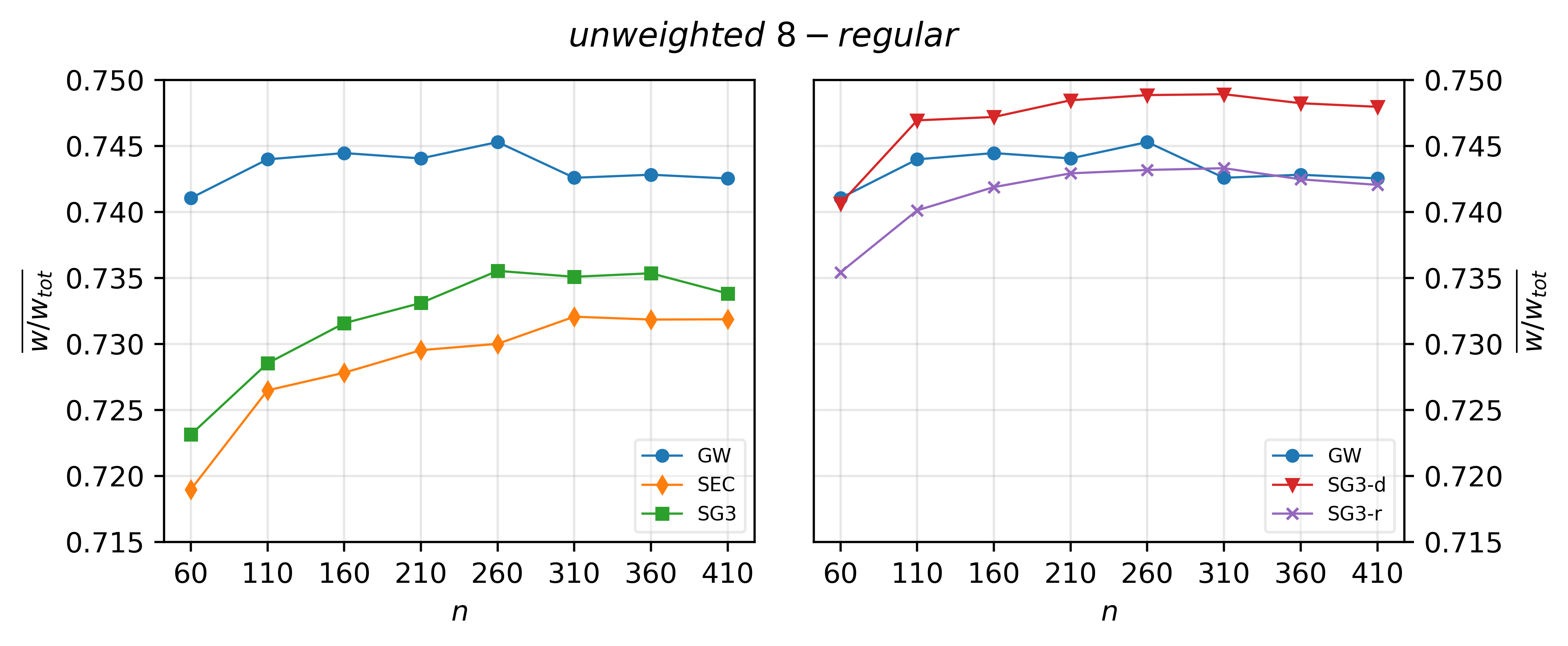}
\caption{\it Average performance of various algorithms for the unweghted $8$-regular graphs. For each $n$, we generate $40$ instances, calculate the cut ratio $w/w_{tot}$ for each instance, and finally take the average over all instances. Left: results from SEC, SG3 and GW are plotted; Right: results from SG3-d and SG3-r and GW results are shown.}\label{fig:unwt-8r}
\end{figure}

We plot the results for weighted and unweighted $8$-regular graphs in Fig.~\ref{fig:wt-8r} and Fig.~\ref{fig:unwt-8r}. For the weighted case, we sample the edge weights independently from the uniform distribution $U[0,1]$.


As expected, the pattern shown in Fig.~\ref{fig:wt-8r} is very similar to the previous case of complete graphs in Fig.~\ref{fig:Complete-Graph}: SEC and SG3 perform roughly the same, and both are worse than GW, while SG3-d and SG3-r surpass GW, especially in the large $n$ region. When the weights are removed, SEC and SG3 deteriorate further. Although SG3-r also becomes worse than GW, SG3-d still keeps its advantage in the whole region. Therefore the general pattern in Fig.~\ref{fig:wt-8r} is roughly maintained in Fig.~\ref{fig:unwt-8r}. This is reasonable: turning off weights makes the edge-oriented algorithms difficult to proceed, but has relatively weaker effects on vertex-oriented algorithms.

Clearly $8$-regular graphs are not sparse enough. So we decrease the degree to $k=3$, and repeat all the calculations for both weighted and unweighted graphs. The results are plotted in Fig.~\ref{fig:wt-3r} and Fig.~\ref{fig:unwt-3r} below.

\begin{figure}[H]
\centering
	\includegraphics[width=0.85\textwidth]{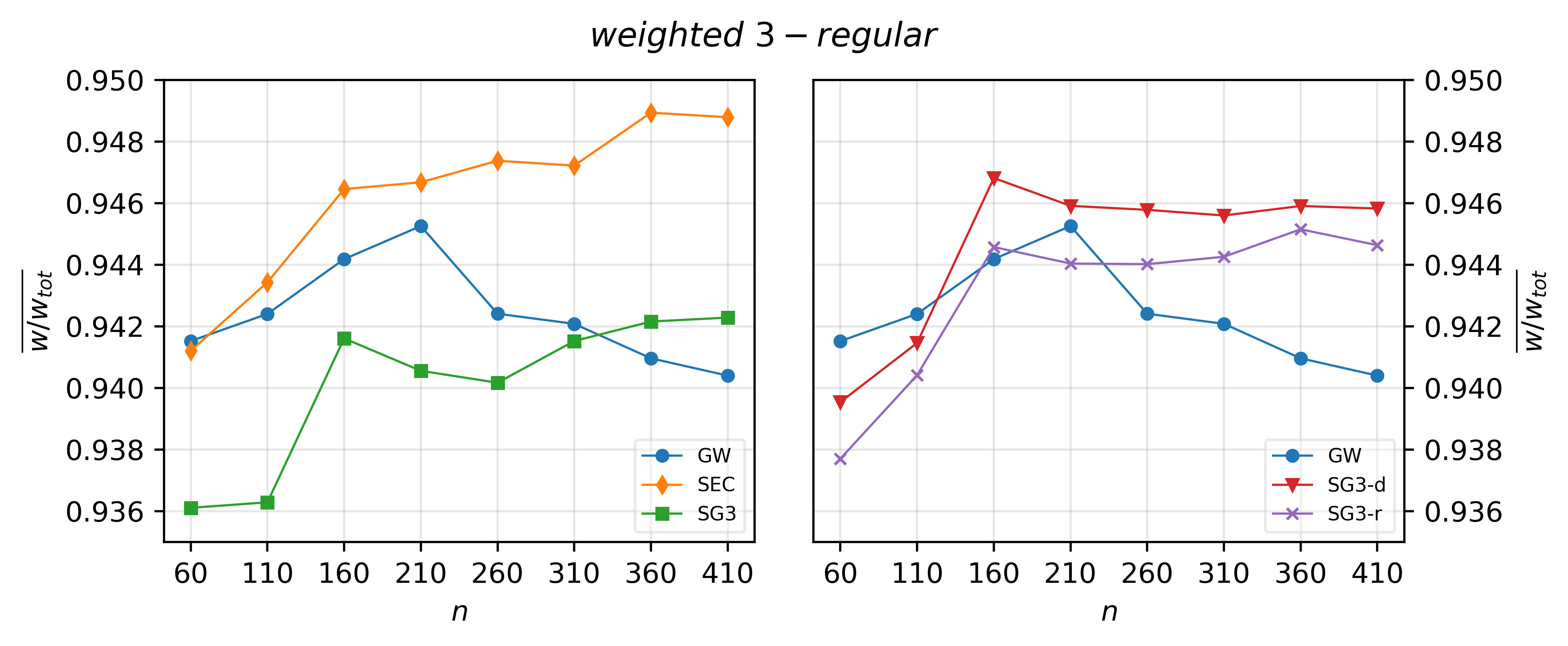}
\caption{\it Average performance of various algorithms for the weighted $3$-regular graphs. For each $n$, we generate $40$ instances with $\{w_{ij}\}$ sampled independently from $U[0,1]$, calculate the cut ratio $w/w_{tot}$ for each instance, and finally take the average over all instances. Left: results from SEC, SG3 and GW are plotted; Right: results from SG3-d and SG3-r and GW results are shown.}\label{fig:wt-3r}
\end{figure}
\begin{figure}[H]
\center
	\includegraphics[width=0.85\textwidth]{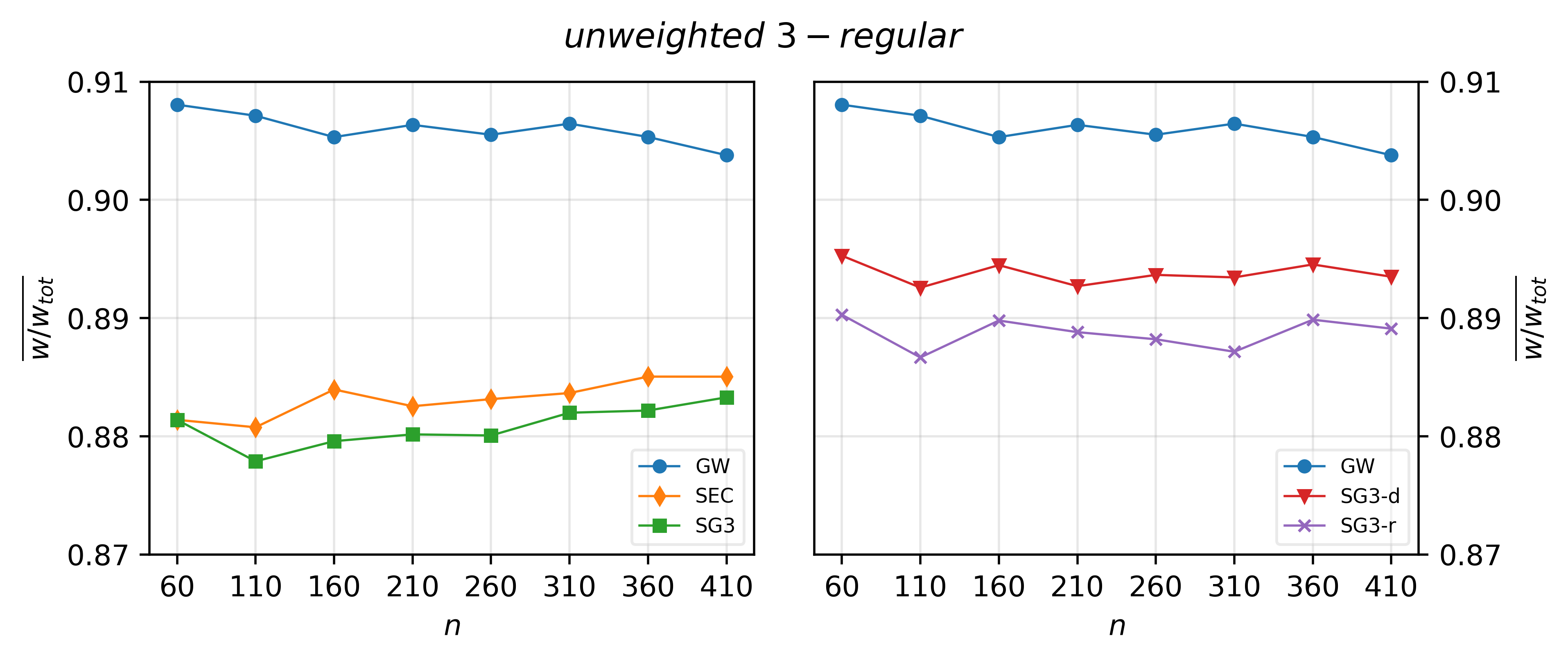}
\caption{\it Average performance of various algorithms for the unweighted $3$-regular graphs. For each $n$, we generate $40$ instances, calculate the cut ratio $w/w_{tot}$ for each instance, and finally take the average over all instances. Left: results from SEC, SG3 and GW are plotted; Right: results from SG3-d and SG3-r and GW results are shown.}\label{fig:unwt-3r}
\end{figure}

First we see dramatic changes for the weighted $3$-regular graphs in Fig.~\ref{fig:wt-3r}: SG3 now catches up with GW when $n$ is large, and SEC dominates over GW in the whole region. Meanwhile, both SG3-d and SG3-r deteriorate and are overtaken by GW when $n$ is small. Most importantly, SEC surpasses SG3-d almost in the whole region, as expected.

However, when the weights are turned off, a disaster occurs, as shown in Fig.~\ref{fig:unwt-3r} : all the four greedy algorithms deteriorate a lot, and fall far below GW. The situation with SEC and SG3 is the worst. Such a deterioration has been noticed in~\cite{ADAPT-Clifford}.


\subsubsection{Erd\"{o}s-R\'{e}nyi graphs}

It will be natural to extend the previous study to more general sparse graphs, those with fixed average degree. As a typical representative, we consider random Erd\"{o}s-R\'{e}nyi~(ER) graphs with fixed edge density, $G(n,p)$. Then the average degree would approximately be $\bar k \approx np$ when $n$ is not too small.

\begin{figure}[H]
\centering
	\includegraphics[width=0.85\textwidth]{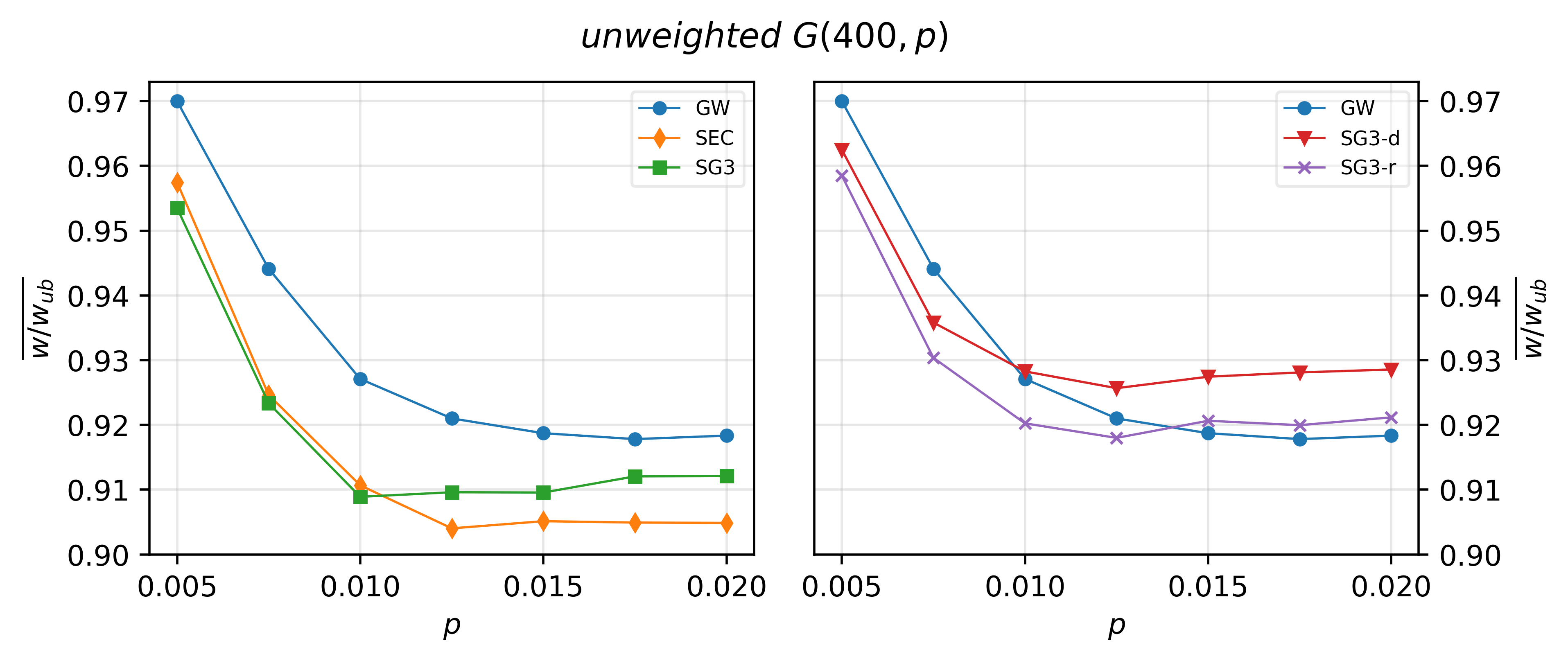}
\caption{\it Average performance of various algorithms for the unweighted ER graphs $G(n,p)$ with fixed size $n=400$. For each $p$, we generate $40$ instances, calculate the cut ratio $w/w_{ub}$ for each instance, and finally take the average over all instances. Left: results from SEC, SG3 and GW are plotted; Right: results from SG3-d and SG3-r and GW results are shown.}\label{fig:unwt-ER-p}
\end{figure}

\begin{figure}[H]
\center
	\includegraphics[width=0.85\textwidth]{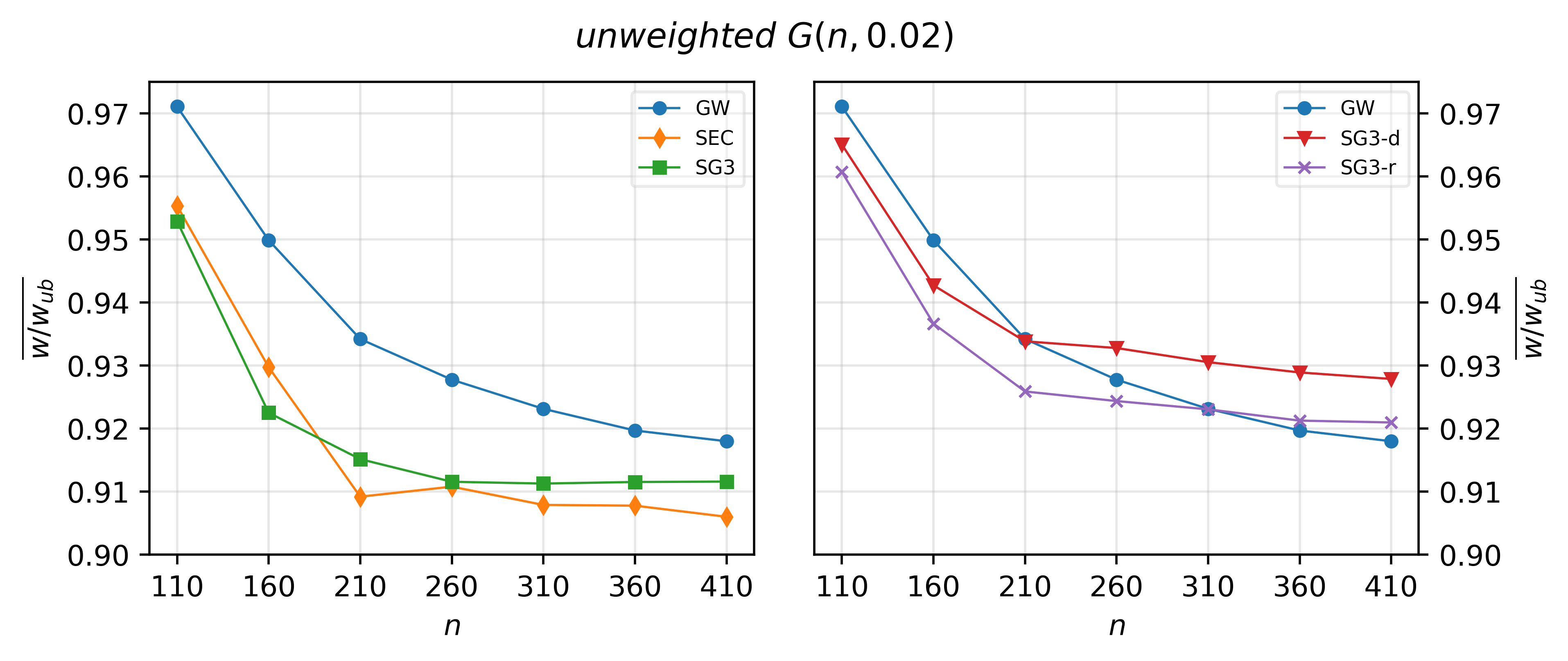}
\caption{\it Average performance of various algorithms for the unweighted ER graphs $G(n,p)$ with fixed density $p=0.02$. For each $n$, we generate $40$ instances, calculate the cut ratio $w/w_{ub}$ for each instance, and finally take the average over all instances. Left: results from SEC, SG3 and GW are plotted; Right: results from SG3-d and SG3-r and GW results are shown.}\label{fig:unwt-ER-n}
\end{figure}

We consider two situations: varying $n$ with fixed $p$, and varying $p$ with fixed $n$. In both cases we make sure that the range of the average degree $\bar k$ covers the interval $[3,8]$. For each parameter set $(n,p)$, we generate $40$ instances, calculate the ratio $w/w_{ub}$ with respect to the SDP upper bound, and finally take the average over all instances. Here we choose the SDP upper bound rather than the total weight as the denominator, because it helps to discriminate various algorithms. We start with the unweighted case, first vary $p$ and then vary $n$, and plot the results in Fig.~\ref{fig:unwt-ER-p} and Fig.~\ref{fig:unwt-ER-n} respectively.


The results in the above two figures confirm the conclusions made in the previous subsection. Namely, when the average degree decreases, performance of vertex-oriented algorithms deteriorates, while that of the edge-oriented algorithms improves. This is clearly seen in the right part of Fig.~\ref{fig:unwt-ER-p}: at large $p$, and thus large $\bar k$, SG3-d is superior to GW; when $p$ and $\bar k$ decrease, SG3-d deteriorates and becomes worse than GW. A kind of ``phase transition'' occurs at $\bar k_c\approx 4$. Similar phase transitions have been studied extensively in the literature, especially in the propositional satisfiability~(SAT) problems~\cite{Monasson-1999}. On the left part of Fig.~\ref{fig:unwt-ER-p}, we find the improvement of SEC as $p$ and $\bar k$ decrease, which exceeds SG3 when $\bar k$ is smaller than $\bar k_c\approx 4$. The results in Fig.~\ref{fig:unwt-ER-n} exhibits almost the same pattern as Fig.~\ref{fig:unwt-ER-p}, where now $n$ is varied instead of $p$. And the phase transition point is again around the critical degree $\bar k_c\approx 4$.

\begin{figure}[H]
\centering
	\includegraphics[width=0.85\textwidth]{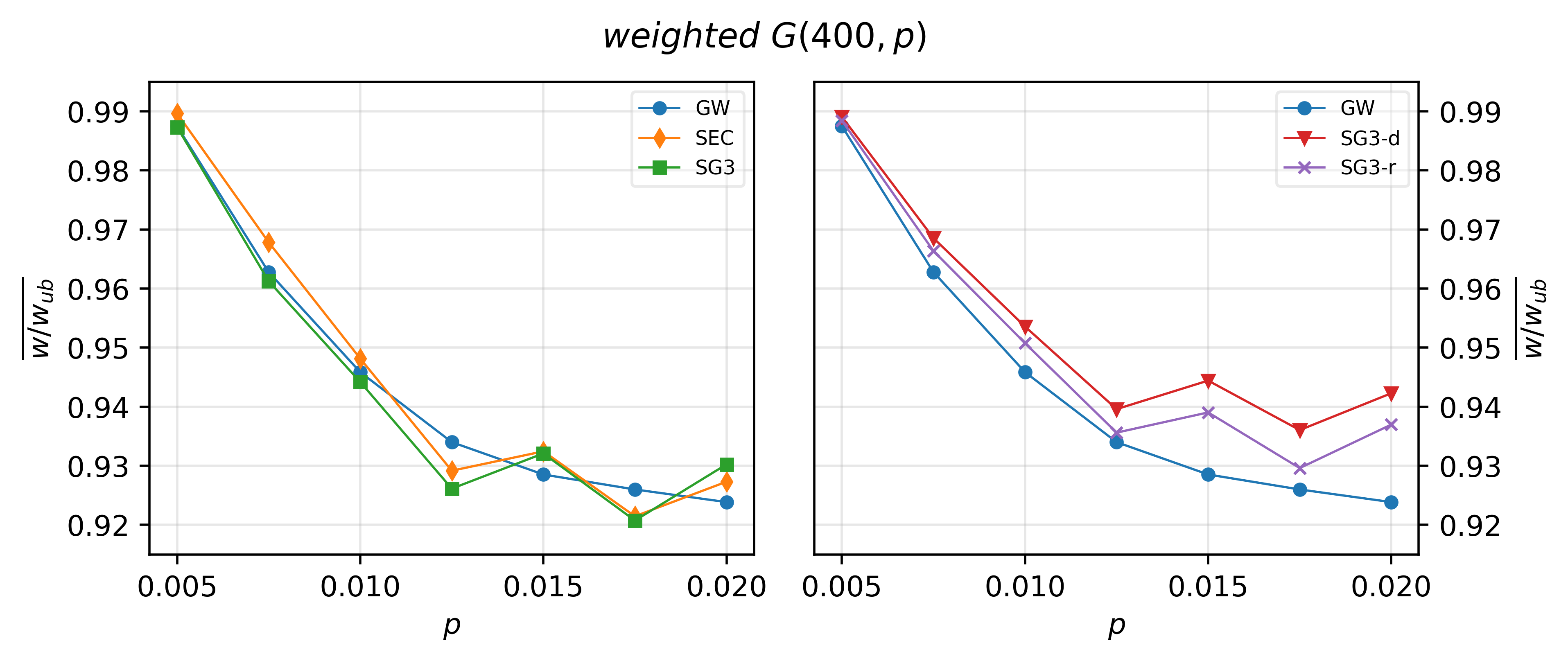}
\caption{\it Average performance of various algorithms for the weighted ER graphs $G(n,p)$ with fixed size $n=400$. For each $p$, we generate $40$ instances with the edge weights $\{w_{ij}\}$ sampled independently from $U[0,1]$, calculate the cut ratio $w/w_{ub}$ for each instance, and finally take the average over all instances. Left: results from SEC, SG3 and GW are plotted; Right: results from SG3-d and SG3-r and GW results are shown.}\label{fig:wt-ER-p}
\end{figure}

\begin{figure}[H]
\center
	\includegraphics[width=0.85\textwidth]{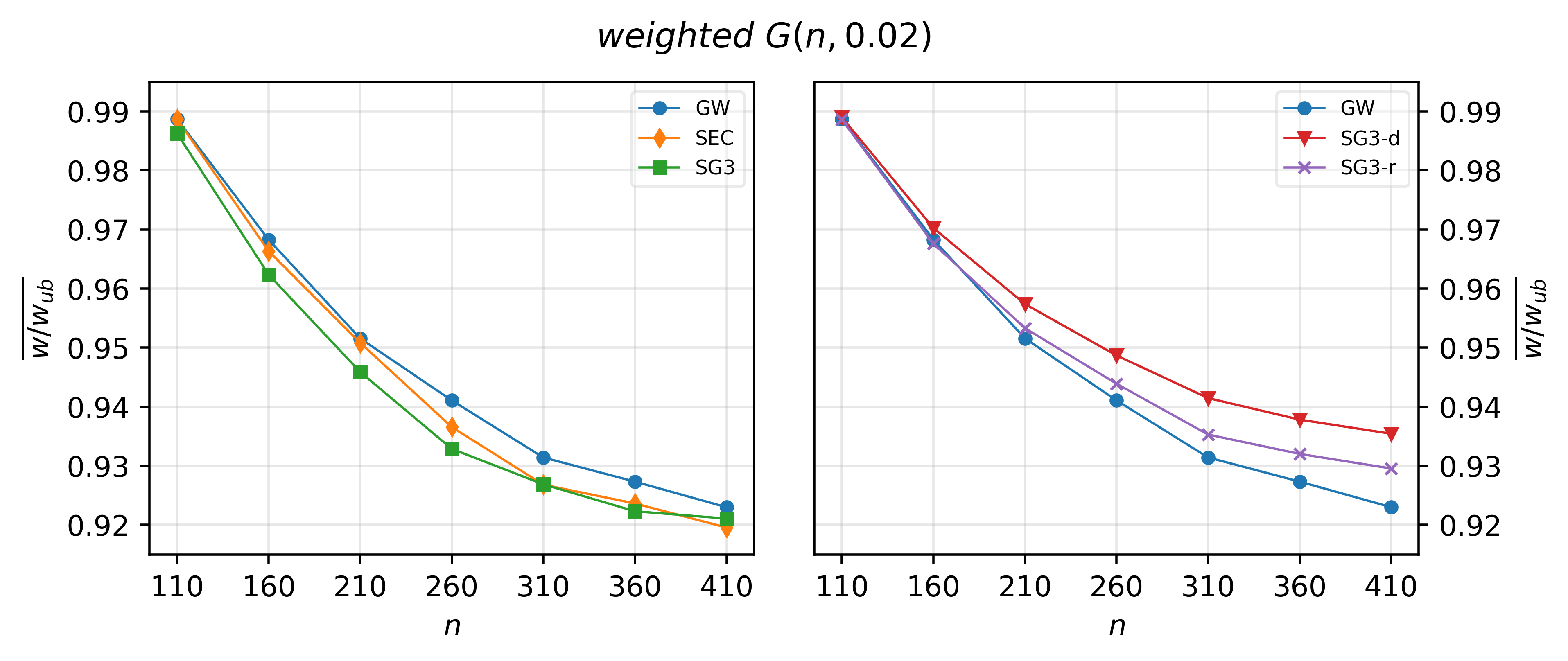}
\caption{\it Average performance of various algorithms for the weighted ER graphs $G(n,p)$ with fixed density $p=0.02$. For each $n$, we generate $40$ instances with the edge weights $\{w_{ij}\}$ sampled independently from $U[0,1]$, calculate the cut ratio $w/w_{ub}$ for each instance, and finally take the average over all instances. Left: results from SEC, SG3 and GW are plotted; Right: results from SG3-d and SG3-r and GW results are shown.}\label{fig:wt-ER-n}
\end{figure}

Next we turn on the edge weights by sampling them independently from $U[0,1]$. The results with varying $p$ and $n$ are shown in Fig.~\ref{fig:wt-ER-p} and Fig.~\ref{fig:wt-ER-n} respectively.


From the left part of Fig.~\ref{fig:wt-ER-p}, we see that the improvement of SEC with decreasing $\bar k$ is enhanced when the edge weights are introduced. SEC now exceeds GW when $\bar k\le 4$. In the right part, the deterioration of SG3-d with decreasing $\bar k$ is now mitigated with the weights turned on. The cross of SG3-d and GW now disappears. We may also compare the results of SEC with SG3-d directly. It turns out that even in the region $\bar k < 4$,  SEC does not strictly surpass SG3-d, as in the $3$-regular graphs. This should not be surprising: a nonzero distribution for the degree helps mitigate the sparsity problem for SG3-d. The results in Fig.~\ref{fig:wt-ER-n} are similar to Fig.~\ref{fig:wt-ER-p}: below $\bar k_c\approx 4$, the results of SEC almost coincides with that of GW, although do not strictly surpass the latter.

In summary, a kind of phase transition is found to occur at the critical average degree $\bar k_c \approx 4$. This is mostly clearly seen from the crossing of the SG3-d and GW curves in the unweighted case, and that of SEC and GW in the weighted case.

\section{Summary}

 A new concept, relation tree, is introduced to represent graph cut. With this, various greedy heuristics for MAX-CUT could be classified as two different classes of tree algorithms: the vertex-oriented Prim class, and the edge-oriented Kruskal class. The SG series of algorithms belong to the Prim class, while the EC series belong to the Kruskal class. We also give an algebraic formulation of this framework, based on the stabilizer formalism in quantum computing. Such an algebraic formulation helps to derive the approximation ratio of various algorithms. It also helps to extend these algorithms to more general cases, especially to the intrinsic quantum problems~\cite{SA-I}.

 The classification base on relation tree is then utilized to improve present algorithms. We show that modifying the initialization of SG3 from the maximum edge to an arbitrary vertex greatly improves the performance. Such an improvement is so significant that on most graphs the modified algorithms, SG3-d and SG3-r, outperform the GW algorithm based on SDP. In particular, in the SK model SG3-d could achieve a regularized energy as low as $-0.70$ at large size, which is nearly $92\%$ of the Parisi value and far below the SDP limit. For a greedy heuristic, such a performance is quite remarkable.

We also use such a classification to assess the performance of different classes of algorithms. It turns out that their performance mainly depend on the average degree of a graph. In general, a phase transition occurs around a critical degree $\bar k_c\approx 4$. Above this degree, Prim-class algorithms such as SG3-d outperforms all others and even GW. Below this degree SG3-d gradually deteriorates. The Kruskal-class algorithms, such as SEC, behave in the opposite way: below the critical degree, SEC could be competitive with GW, and could even surpass SG3-d if the degree distribution is sharp; when the average degree increases above $\bar k_c$, SEC quickly deteriorates. For SG3-d, the phase transition is significant in the unweighted case, which would be smoothed out when the edge weights are included. And for SEC, the phase transition is more clearly seen in the weighted case, and less significant when the weights are turned off.



\end{document}